\documentclass[journal=jacsat,manuscript=article]{achemso}

\usepackage[version=3]{mhchem} 
\usepackage{chemformula}
\usepackage[dvipsnames]{xcolor}
\usepackage{amssymb}

\newcommand{\fb}[1]{\textcolor{black}{#1}}

\author{Jacopo Vialetto}
\affiliation{Department of Chemistry, University of Florence, via della Lastruccia 3, Sesto Fiorentino, I-50019 Firenze, Italy}
\alsoaffiliation{Consorzio interuniversitario per lo sviluppo dei Sistemi a Grande Interfase (CSGI), via della Lastruccia 3, 50019 Sesto Fiorentino (FI), Italy}
\email{jacopo.vialetto@unifi.it}

\author{Francesco Brasili}
\affiliation{Institute for Complex Systems, National Research Council, Piazzale Aldo Moro 5, 00185, Roma, Italy}
\alsoaffiliation{Department of Physics, Sapienza University of Rome, Piazzale Aldo Moro 5, 00185 Roma, Italy}

\author{Letizia Tavagnacco}
\affiliation{Institute for Complex Systems, National Research Council, Piazzale Aldo Moro 5, 00185, Roma, Italy}
\alsoaffiliation{Department of Physics, Sapienza University of Rome, Piazzale Aldo Moro 5, 00185 Roma, Italy}

\author{Gavino Bassu}
\affiliation{Department of Chemistry, University of Florence, via della Lastruccia 3, Sesto Fiorentino, I-50019 Firenze, Italy}
\alsoaffiliation{Consorzio interuniversitario per lo sviluppo dei Sistemi a Grande Interfase (CSGI), via della Lastruccia 3, 50019 Sesto Fiorentino (FI), Italy}

\author{Elena Buratti}
\affiliation{Department of Environmental and Prevention Sciences, University of Ferrara, Via L. Borsari, 46, 44121 Ferrara, Italy}

\author{Stephen King}
\affiliation{ISIS Neutron and Muon Source, Rutherford Appleton Laboratory, Oxfordshire, OX11 0QX, UK}

\author{Emanuela Zaccarelli}
\affiliation{Institute for Complex Systems, National Research Council, Piazzale Aldo Moro 5, 00185, Roma, Italy}
\alsoaffiliation{Department of Physics, Sapienza University of Rome, Piazzale Aldo Moro 5, 00185 Roma, Italy}
\email{emanuela.zaccarelli@cnr.it}

\author{Marco Laurati}
\affiliation{Department of Chemistry, University of Florence, via della Lastruccia 3, Sesto Fiorentino, I-50019 Firenze, Italy}
\alsoaffiliation{Consorzio interuniversitario per lo sviluppo dei Sistemi a Grande Interfase (CSGI), via della Lastruccia 3, 50019 Sesto Fiorentino (FI), Italy}
\email{marco.laurati@unifi.it}

\title[An \textsf{achemso} demo]
  {Heterogeneous collapse in thermoresponsive copolymer microgels varying molar composition}

\begin{document}
\newcommand{\ml}[1]{\textcolor{black}{#1}}
\newcommand{\jv}[1]{\textcolor{black}{#1}}
\begin{tocentry}





\includegraphics[width=\textwidth]{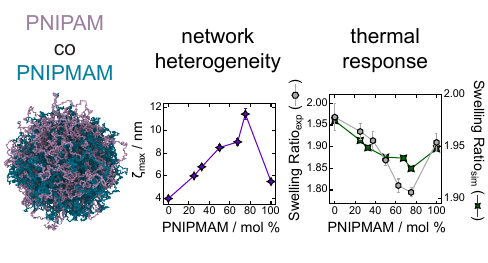}

\end{tocentry}


\begin{abstract}
Understanding the internal architecture of copolymer microgels is crucial for establishing how nanoscale polymer organization controls their stimuli-responsive behavior.
Here we focus on thermoresponsive P(N-isopropylacrylamide-\textit{co}-N-isopropyl-methacrylamide), P(NIPAM-\textit{co}-NIPMAM), microgels with varying mole fraction of the components, synthesized via radical precipitation polymerization, and we demonstrate that changes in their volume phase transition and equilibrium swelling are governed by composition-dependent internal heterogeneity.
Comparison between small-angle neutron scattering (SANS) with isotopic labeling and monomer-resolved simulations show a block-like monomer distribution of the two components.
SANS analysis reveals a universal maximum in the polymer mesh correlation length near the transition, evidencing coexistence of collapsed NIPAM-rich and swollen domains.
The correlation length increases with increasing NIPMAM content, with a maximum for 75 mol \% NIPMAM, implying sparse collapsed regions within the network and thus a large degree of heterogeneity induced by the presence of an increasingly large fraction of intercalated, non-collapsing PNIPMAM.
The maximum heterogeneity correlates with the equilibrium swelling ratio, indicating that collapsed microgels retain a structural memory of the transition and present a less-compliant structure in response to temperature variations.
Overall, these insights highlight a complex effect of the block-like monomer distribution on the responsive properties of copolymer microgels with different compositions, thus providing a design rule for tailoring responsive colloids for functional soft materials.

\end{abstract}




\section{Introduction}

Stimuli-responsive colloids are an important class of materials with significant interest for fundamental studies in condensed matter physics, and for multiple potential applications ranging from drug delivery to additive manufacturing or surface coatings. \cite{Karg2019,Highley2019,Rey2020,Xie2025} In particular, colloids made of crosslinked polymeric networks (i.e. microgels) offer a platform potentially responsive to multiple stimuli (pH, temperature, light, specific molecules, etc.) that can be tailored to the users needs by choosing the network components.
The most studied systems include the thermoresponsive polymer poly(N-isopropylacrylamide) (PNIPAM). PNIPAM microgels undergo a volume phase transition (VPT) at a critical temperature ($T_{VPT} \simeq 32 ^\circ C$), from a swollen to a collapsed particle due to a decrease of solvent affinity of the polymer. \cite{Plamper2017}
Superior control over microgel's responsiveness to various stimuli can be achieved by incorporating a second polymer during the particle synthesis. \cite{Hertle2013} Copolymerization of PNIPAM microgels provides pH or light responsiveness, \cite{Kratz2000,del2021two,Phua2016,Hu2022} or it allows tuning $T_{VPT}$ to meet the required functions.
In particular, comonomers such as N-isopropylmethacrylamide (NIPMAM), \cite{Keerl2009,Clarke2013} N-n-propylacrylamide (NnPAM) \cite{Wiehemeier2019} or oligo(ethylene glycol) methyl ether methacrylate (OEGMA) \cite{Clara-Rahola2012,Wellert2015,Motlaq2018,Agnihotri2020,Vialetto2026} allow increasing pure PNIPAM $T_{VPT}$ to match the body temperature, of interest for applications such as drug delivery and tunable cell adhesion, where specific functions can be tuned in physiological conditions. \cite{Kuroki2012}

An important aspect controlling microgel responsivity and interactions is the architecture of the polymeric network. 
Different responses can be achieved when two (or more) polymers are spatially separated, one in the core and one in the shell of the particle, as in core-shell microgels, or if the microgels are synthesized with statistical copolymerization. In core-shell microgels two distinct $T_{VPT}$ are present, corresponding to the collapsing temperature of each of the homopolymer region of the particle. \cite{Berndt2006,Keerl2008,Hu2010}
If instead the two monomers are added together during particle synthesis (i.e. one-pot synthesis), one single $T_{VPT}$ is typically obtained. \cite{Balaceanu2013,Wiehemeier2019}
In such cases, the relative distribution of the components within the polymeric network will depend on the reaction constants \cite{Duracher1999,Hoare2006} and on the interactions between the polymer chains.
Although this will ultimately dictate the overall network response and is fundamental in order to predict material properties, to date little information is known on the detailed internal copolymer distribution as most techniques do not provide access to this information.
Additionally, even more basic properties, such as the dependence of $T_{VPT}$ on composition, are not straightforward. For example, in PNIPAM-PNIPMAM microgels $T_{VPT}$ changes linearly with PNIPMAM content. \cite{Clarke2013} Instead, for poly(N-vinylcaprolactam) PVCL-PNIPMAM particles, Balaceanu and coworkers showed that $T_{VPT}$ increases non-linearly from 34°C to 45°C upon increasing the PNIPMAM fraction, with the behavior of copolymer microgels made with ratios higher or lower than 1 dominated by VCL or NIPMAM, respectively. \cite{Balaceanu2013}
This indicates that the specific interactions between the monomers might yield significantly different results and that a precise characterization of these systems is required in order to expand their applicability in various fields.

Here, we focus on the well-known P(NIPAM-\textit{co}-NIPMAM) system. PNIPMAM, displaying a $T_{VPT}\simeq 44^\circ C$, \cite{Wedel2016} allows modulation of the transition temperature in copolymer networks in the physiological range. \cite{Djokp2001,Clarke2013} A previous work analyzed microgels at equimolar composition, observing that the $T_{VPT}$ of the copolymer network is $38^\circ C$ and thus in between that of the corresponding homopolymers. \cite{Keerl2009} Recently, by combining small-angle neutron scattering (SANS) and nuclear magnetic resonance (NMR) experiments with numerical simulations, we probed the swelling behavior of the microgels at equimolar composition across $T_{VPT}$ unveiling the formation of local domain structures (named block topology) within the network. These domains are composed of repeating units of the same type (either NIPAM or NIPMAM), which preferentially cluster together rather than being randomly and homogeneously distributed. \cite{Tavagnacco2025} These results indicate that the internal conformation of copolymer microgels is not necessarily a simple statistical distribution, even in a one-pot synthesis, and that such compositional heterogeneity can be revealed only by probing the inner structure of the particles.
Taking a step further, here we investigate P(NIPAM-\textit{co}-NIPMAM) microgels at varying composition by combining dynamic light scattering (DLS), \jv{differential scanning calorimetry (DSC)}, SANS and numerical simulations to characterize their internal structure and thermal responsivity, and we evidence that such internal heterogeneity is composition dependent, reaching a maximum for 75 mol \% PNIPMAM. The heterogeneity at the transition temperature are reflected in variations in the equilibrium swelling ratio, suggesting that the collapsed microgels preserve a structural signature of the transition, which modulates their temperature response and deswelling.

\section{Materials and methods}

\subsection{Reagents}
N-isopropylacrylamide (NIPAM, $\text{MW}=113.16$ Da, Sigma, 97\% purity), N-isopropylmethacrylamide (NIPMAM, $\text{MW}=127.18$ Da, Sigma, 97\% purity), N,N'-methylenebisacrylamide (BIS, $\text{MW}=154.17$ Da, Sigma, 99\% purity), Sodium dodecyl sulfate (SDS, $\text{MW}=288.38$ Da, Sigma, 99\% purity) potassium persulfate (KPS, $\text{MW}=270.32$ Da, Sigma, 99\% purity) \jv{and deuterium oxide (Sigma, 99.9 atom \% D)}.
NIPAM and NIPMAM were purified by recrystallization in hexane. Water used for all experiments is bi-distilled Milli-Q.

\subsection{Microgel synthesis}
The microgels used in this work were synthesized by radical precipitation polymerization. 
We used 5\% BIS as crosslinker and 1\% KPS as initiatior. The reaction was carried out in a solution with 4 mM SDS surfactant. We used different molar percentage of the NIPAM and NIPMAM monomers, as reported in Table \ref{tbl:synthesis}.
Briefly, the monomers, crosslinker and surfactant were dissolved in 26.5 mL of deionized water to yield final concentrations of 160 mM, 8 mM, and 4 mM, respectively.
The solution was loaded in a 50 mL two-necked reactor equipped with a condenser and a magnetic stirrer, and then bubbled under a nitrogen stream for 1 hour at room temperature. Separately, the initiator was dissolved in degassed water at the concentration of 36.9 mM. The temperature in the reactor was raised to 70°C using an oil bath and then the polymerization reaction was initiated by adding 1.2 mL of the initiator solution at the rate of 1 mL/min. The final concentration of the initiator in the reaction solution was 1.6 mM. The reaction was carried out for 5 hours. The resulting microgels were purified by dialysis against ultrapure water using a cellulose membrane (6–8 kDa MWCO, Sigma) for two weeks, with the water replaced twice daily. The microgel solutions were then freeze-dried and stored in the dark at 4°C.

For the synthesis of microgels with deuterated NIPAM, P(D-NIPAM-co-H-NIPMAM) with 25\% NIPAM and 75\% NIPMAM monomers, we used 8 mol\% H-NIPAM and 17 mol\% NIPAM-d$_{10}$, while keeping the same molar concentrations of the other reagents and synthesis procedure. 
The PNIPAM composition is chosen in order to match the scattering length density with that of the solvent. NIPAM-d$_{10}$ monomers were synthesized at the J\"ulich Centre for Neutron Scattering according to the procedure reported in ref.~\citenum{Tavagnacco2025}. 

\begin{table}
  \caption{Microgels used in this work}
  \label{tbl:synthesis}
  \begin{tabular}{llllll}
    \hline
    Microgel & NIPAM mol \% & NIPMAM mol \% & R$_h$ (20°C) [nm] & $\frac{R_h (25^\circ C)}{R_h (T_{max}~^\circ C)}$ & VPTT [°C] \\ 
    \hline
M$_{100-0}$     & 100  & 0    & 58.5 $\pm$ 0.4 & 1.97 $\pm$ 0.03 & 33.1 $\pm$ 0.1 \\  
M$_{75-25}$     & 75   & 25   & 72.5 $\pm$ 0.2 & 1.93 $\pm$ 0.02 & 35.0 $\pm$ 0.1 \\    
M$_{62.5-37.5}$ & 62.5 & 37.5 & 81.0 $\pm$ 0.7 & 1.91 $\pm$ 0.02 & 36.1 $\pm$ 0.1 \\    
M$_{50-50}$     & 50   & 50   & 79.4 $\pm$ 0.3 & 1.87 $\pm$ 0.01 & 37.9 $\pm$ 0.1 \\    
M$_{37.5-62.5}$ & 37.5 & 62.5 & 68.4 $\pm$ 0.8 & 1.81 $\pm$ 0.02 & 39.2 $\pm$ 0.1 \\    
M$_{25-75}$     & 25   & 75   & 96.5 $\pm$ 0.5 & 1.79 $\pm$ 0.01 & 40.9 $\pm$ 0.1 \\    
M$_{0-100}$     & 0    & 100  & 95.5 $\pm$ 0.5 & 1.91 $\pm$ 0.02 & 44.4 $\pm$ 0.1 \\    
    \hline    
  \end{tabular}
\end{table}

\subsection{Experimental methods}

\noindent \textit{Dynamic light scattering.} Dynamic light scattering (DLS) experiments were performed using a Zetasizer PRO Red Label (Malvern, UK). The microgels were redispersed in Milli-Q water at a concentration of 0.01 wt\% in order to analyze them in the dilute regime.
The temperature was varied from 25 to 50°C or 65°C with 1°C steps and 5 minutes equilibration time. 
\fb{For each temperature we recorded three consecutive intensity autocorrelation functions, each averaged over 20 runs.}
\fb{We then extrapolated decay times by cumulant analysis of the correlograms, and derived the distribution of diffusion coefficients $D$ of the microgels.
Intensity-weighted distributions of hydrodynamic radius $R_h$ were obtained using the Stokes-Einstein relationship $R_H = k_BT/6\pi\eta D$, where $k_BT$ is the thermal energy and $\eta$ the water viscosity.}
\fb{The temperature-responsive change in hydrodynamic radius $R_h(T)$ was fitted to the phenomenological function~\cite{DelMonte2021,Ballin2025}}
\begin{equation}\label{eq:DLS}
\fb{R_h(T)=R_0-\Delta R\,\tanh{\left[ s\left(T-T_{VPT} \right)\right]}+A\left(T-T_{VPT}\right)+B\left(T-T_{VPT}\right)^2\quad ,}
\end{equation}
\fb{where $R_0 = \frac{1}{2}(R_\text{max}+R_\text{min})$ and $\Delta R = \frac{1}{2}(R_\text{max}-R_\text{min})$, with $R_\text{max}$ and $R_\text{min}$ the plateau values of $R_h(T)$ at low (fully swollen microgels) and high (fully collapsed microgels) temperatures, respectively, $T_{VPT}$ is the critical temperature of the VPT, $s$ is the sharpness parameter, whose inverse quantifies the temperature range over which the VPT occurs. The linear and quadratic terms,} \jv{whose fitting coefficients are A and B, respectively,} \fb{are phenomenological corrections that allow to accurately reproduce the non-flat baselines of the experimental trends outside the temperature range of the VPT, occurring in particular at low temperatures.} 
\fb{In this way, the function is able to accurately capture the behavior of $R_h(T)$ in the whole investigated temperature range, avoiding spurious effects on the physical parameters such as $T_{VPT}$ and $s$.}

\fb{We also defined the experimental swelling ratio $SR_{exp}$ as the ratio between the hydrodynamic radii $R_h$(25°C) and $R_h$($T_{max}$), in swollen and collapsed conditions, respectively:}
\begin{equation}\label{eq:SR}
SR_{exp} = \frac{R_h (25\text{°C})}{R_h (T_{max})}
\end{equation}
\fb{In order to account for fully collapsed particles, we selected $T_{max}$ well-above $T_{VPT}$ for all the analyzed samples. Specifically, $T_{max}=50$°C up to 37.5 mol \% NIPMAM and $T_{max}=65$°C for samples with larger NIPMAM content.}

\noindent \jv{\textit{Differential scanning calorimetry (DSC).}
Thermal analyses on microgels were performed with a Perkin Elmer Pyris Diamond DSC equipped with Intracooler III as cooling system and a DSC 8000 Perkin Elmer differential scanning calorimeter equipped with Intracooler II as cooling system. About 10 mg of microgel dispersions at $C_w$ = 10 wt\% was analysed under nitrogen atmosphere (20 mL/min) in hermetic sealed steel pans. The measurements were carried out by cooling the system from 20 to 0 °C, holding the temperature for 1 min, then heating it to 80 °C, holding the temperature for 1 min and finally by cooling it again to 20 °C. Each cooling and heating step was carried out with a scanning rate of 5 °C/min. Analysis of the thermograms was performed with \textit{Pyris 7} software.}

\noindent \textit{Small-angle neutron scattering (SANS).}
Mixtures of microgels (0.1 w/w \%) 
were prepared by diluting stock solutions of freeze-dried microgels in D$_2$O. The obtained samples were then mixed at room temperature in an orbital shaker overnight. SANS measurements were performed at SANS2D (ISIS, Didcot, UK) and SANS-I (PSI, Villigen, Switzerland) beamlines. 
At SANS2D we used a white beam of wavelength $ 1.75<\lambda<12.5$ \AA~, an incident collimation of 12 m and two detectors placed 5 and 12 m from the sample. These settings gave a simultaneous wave vector range of 0.0015 $\text{\AA}^{-1} < Q < 0.50\text{ \AA}^{-1}$. While at SANS-I we used a constant $\lambda$ = 8 \AA~for three sample-to-detector distances of 1.6 m, 4.5 m and 18 m, \jv{and collimations of 3, 4.5 and 18 m, respectively}. The combination of the three configurations gave a wave vector range of 0.0018 $\text{\AA}^{-1} < Q < 0.25\text{ \AA}^{-1}$.

All samples were measured at different temperatures between 20°C and 50°C in quartz cells with a path length of 2~mm (Hellma GmbH \& Co., Mullheim, Germany). After reaching the desired temperature, the samples were equilibrated for at least 5 min. All scattering data were normalized for the sample transmission and background corrected using a quartz cell filled with D$_2$O. All data were analyzed within the SASView package \jv{(version 5.0.6, 10.5281/zenodo.7581379)} using user-written functions. 
\jv{The SANS scattered intensities of the microgel suspensions can be expressed as:
\begin{equation}
I(q)=  \phi V(\Delta \rho)^{2} P(q)S(q)+\mathrm{bkg}
\label{eq:sans_model}
\end{equation}
where $\phi$ is the microgel volume fraction, $V$ the microgel volume, $\Delta \rho=\rho_{1}-\rho_{2}$ the scattering length density difference between the microgels ($\rho_1$) and D$_2$O ($\rho_2$), $P(Q)$ the particle form factor, S(Q) the interparticle structure factor and $\mathrm{bkg}$ the background. 
}

$P(Q)$ was described using the \jv{well-known} fuzzy sphere model from Stieger et al. \cite{Stieger2004}:
\begin{equation}\label{eq:Pfuzzy}
P(q)= A_1\left[ \frac{3[sin(qR) - qRcos(qR)]}{(qR)^3} \cdot \exp{\left( - \frac{(\sigma q)^2}{2} \right)} \right]^2 + A_{2}\frac{1}{1 + (q\zeta)^2}
\end{equation}
where the first term models the fuzzy sphere contribution, with $R$ the particle radius and $\sigma$ the fuzziness parameter, while the second term describes the polymer network scattering in the region at high Q, with $\zeta$ the correlation length of the polymer mesh. The contributions of the two terms \jv{are} weighted by the amplitudes $A_{1}$ and $A_{2}$.\\
\ml{For samples with low effective microgel concentration (e.g. M$_{25-75}$ in Figure \ref{fig2}), the structure factor contribution is negligible and therefore $S(Q)=1$. Instead, for samples with a noticeable structure factor (for instance M$_{75-25}$ in Figure S11 of the SI), we modeled the $S(Q)$ contribution using a hard-sphere structure factor calculated using the Percus-Yevick closure relationship and containing two parameters, an effective hard-sphere radius $R_{HS}$ and an effective hard-sphere packing fraction $\varphi_{HS}$. We should note that our focus is on extracting changes in the particle morphology associated with the VPTT and as a function of the microgel composition. For this reason we decided to use a simple hard-sphere model to describe the shape of the experimental S(Q) rather than a more accurate description using the Hertzian interaction potential, as done in previous work. \cite{Bassu2024}}
For the fits, the model function was convoluted with the following experimental smearing function:
\begin{equation} \label{eq:smear}
    I(q)= \int P(q-q') \left( \frac{1}{2\pi \sigma_{q'}^{2}} \right)^{1/2} \exp{ \left[ - \frac{q'^{2}}{2 \sigma_{q'}^{2}}  \right]}   \text{d}q'
\end{equation}
where $\sigma_{q}$ is the standard deviation of the $q$~resolution, which contains the detector resolution and the beam wavelength spread contributions\cite{pedersen1993}. In addition, sample polydispersity was included considering a Schulz distribution of the radius.

\subsection{Numerical simulations}
We perform monomer-resolved simulations of copolymer microgels obtained following an \emph{in silico} synthesis protocol~\cite{gnan2017silico} which provides a realistic description of the disordered network and the fuzzy sphere structure~\cite{ninarello2019modeling}. We employ copolymer microgels with a \emph{block} topology, where each polymer chain included between two cross-linkers is entirely assigned to a PNIPAM or PNIPMAM homopolymeric segment. Indeed, this topology was shown to be the most accurate to reproduce the experimental data~\cite{Tavagnacco2025}. Microgels are composed by $N\sim42000$ beads. NIPAM and NIPMAM repeating units are described by bivalent patchy particles, while cross-linkers have four attractive patches. The experimental conditions are reproduced by setting the crosslinker concentration to $c$ = 5.0\% and by using copolymer ratios from 0 to 100.

Beads composing the polymer network interact with the Kremer-Grest potential~\cite{grest1986molecular}, thus, experiencing a steric repulsion modelled by the Weeks-Chandler-Anderson (WCA) potential:
\begin{equation}\label{WCA}
  V_\text{WCA}(r)=\begin{cases} 4\varepsilon \left[\left(\frac{\sigma}{r}\right)^{12}-\left(\frac{\sigma}{r}\right)^6\right]+\varepsilon & \qquad \textrm{if}\quad r\leq2^{\frac{1}{6}}\sigma \\
  0 & \qquad \textrm{if}\quad r>2^{\frac{1}{6}}\sigma \\
  \end{cases}
\end{equation}
where $\epsilon$ and $\sigma$ are the energy and length units, respectively. In addition, bonded beads interact via the finitely extensible nonlinear elastic potential (FENE):
\begin{equation}\label{FENE}
  V_\text{FENE}(r)=-\varepsilon k_F R_0^2 \,\log \left[ 1- \left( \frac{r}{R_0\sigma} \right)^2 \right] \qquad r<R_0\sigma
\end{equation}
where $R_0\sigma$ is the maximum bond distance and $k_F$ is a stiffness parameter that determines the bond rigidity. The covalent nature of the network is accounted for by excluding that bonds can break during the simulation.

The solvent effect is implicitly included by using an effective potential which reproduces the change in the polymer-solvent affinity by raising temperature:
\begin{equation}\label{alpha}
  V_{\alpha}(r)=\begin{cases} -\varepsilon\alpha & \qquad \textrm{if}\quad r\leq2^{\frac{1}{6}}\sigma \\
  \frac{1}{2}\alpha\varepsilon \left\{ \cos \left[ \gamma \left( \frac{r}{\sigma} \right)^2 + \beta \right] -1 \right\} & \qquad \textrm{if}\quad 2^{\frac{1}{6}}\sigma < r \leq R_0\sigma \\
  0 & \qquad \textrm{if}\quad r > R_0\sigma
  \end{cases}
\end{equation}
where $\alpha$ is the solvophobicity parameter corresponding to the effective temperature. The swelling curve of the microgels is obtaining by changing the value of $\alpha$ from good ($\alpha=0$) to bad ($\alpha=1.5$) solvent conditions. Here $\gamma = \pi \left (\frac{9}{4} - 2^\frac{1}{3} \right)^{-1}$ and $\beta=2\pi-\frac{9}{4}\gamma$ are constants defining the functional shape of the potential~\cite{soddemann2001generic}.

The bead size is set to $\sigma$ = 1 for NIPAM and $\sigma$ = 1.04 for NIPMAM to mimic their steric volume difference. Furthermore, the different volume phase transition temperature for PNIPAM and PNIPMAM microgels is modelled by using a different value of $\alpha$ for each monomer, based on the relation between $\alpha$ and temperature determined for PNIPAM microgels in Ref.~\citenum{gnan2017silico}. For mixed interactions occurring between NIPAM and NIPMAM beads the average value of $\alpha$ is used. The value of $\alpha$ reported in the manuscript always refers to the PNIPAM $\alpha$-value for convenience.

Simulations are carried out with the LAMMPS package~\cite{plimpton1995fast} in a cubic box with side L = 200 $\sigma$ at a fixed temperature $\frac{k_BT}{\epsilon}=1.0$. Equations of motion are integrated through a Nos\'{e}–Hoover thermostat in the constant NVT ensemble with an integration time-step $\Delta t = 0.002\tau$, where $\tau=\sqrt{\frac{m\sigma^2}{\varepsilon}}$ is the reduced time unit.  The equilibration of each system is carried out for 1$\cdot$10$^6\tau$, followed by a production run of 1$\cdot$10$^7\tau$.

To directly compare numerical results to the experimental measurements, we computed the microgel form factor $P(q)$ and hydrodynamic radius $R_h$. $P(q)$ is given by
\begin{equation}\label{FF}
  P(q)=\frac{1}{N}\sum_{ij}\langle exp (-i\vec{q}\cdot\vec{r}_{ij}) \rangle
\end{equation}
where $q$ is the wavevector, $r_{ij}$ is the distance between the monomers $i$ and $j$ and the calculation is averaged over independent configurations. The hydrodynamic radius $R_h$ is calculated as:\cite{hubbard1993,del2021two}
\begin{equation}
    R_{sim} 
 =2\left[ \int_0^\infty \frac{1}{\sqrt{(a^2+\theta)(b^2+\theta)(c^2+\theta)}} d\theta \right]^{-1}
\end{equation}
where the microgel is approximated as an effective ellipsoid with semiaxes $a$, $b$ and $c$.
\fb{To compare the swelling curves of $R_{sim}$ to the experimental ones, we normalized each value of $R_{sim}$ to that of the same microgel in the fully swollen state ($\alpha=0$ or $T=25$°C). The values of $\alpha$ are mapped to the experimental temperatures based on the relationship reported in ref.~\citenum{Tavagnacco2025}, which was optimized to reproduce the temperature-dependent evolution of the experimental form factors of the microgel M$_{50-50}$. However, focusing on the 50:50 copolymer composition, that calibration assumes equal contributions of NIPAM and NIPMAM to the effective temperature $\alpha$. Here the different compositions of the co-polymer microgels need to be taken into account. We therefore corrected the mapping relation by introducing the relation $\alpha_{f}=\alpha_{0.5} - (f-0.5)\Delta\alpha$, where $\Delta\alpha =\alpha_\text{NIPAM}-\alpha_\text{NIPMAM}$ is the difference between the values of the solvophobicity parameter of the two monomers and $f$ is the molar fraction of NIPMAM.}


\section{Results and discussion}
\begin{figure}[t!]
\centering
\includegraphics[scale=1]{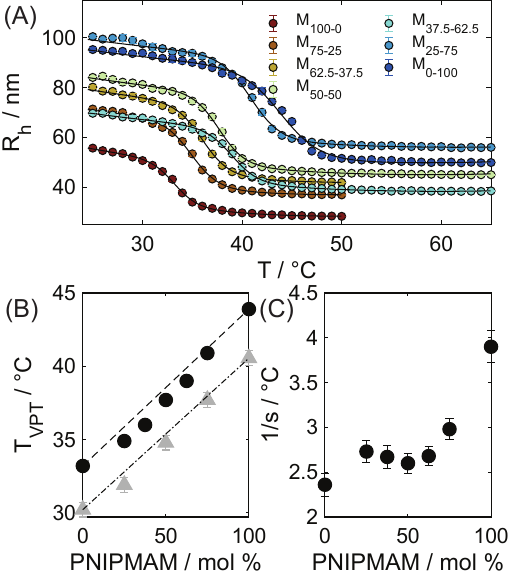}
\caption{\small \jv{PNIPAM-PNIPMAM microgels: experiments. (A) Microgel hydrodynamic radii as a function of temperature. Error bars indicate the standard deviation of 3 measurements. Black lines are fits with eq. \ref{eq:DLS}. (B) $T_{VPT}$ as a function of PNIPMAM composition as obtained from DLS (black circles) and DSC (gray triangles) analysis. In some cases the error bounds are smaller than the symbol size. The dashed and dashed-dotted lines indicates the expected $T_{VPT}$ in case of a linear dependence with the copolymer content for DSL and DSC experiments, respectively. (D) $1/s$ parameter indicating the VPT width.}}
\label{fig1}
\end{figure}

The microgels investigated in this work are synthesized by radical precipitation polymerization, mixing the two monomers NIPAM and NIPMAM in the desired amounts and in the presence of SDS surfactant. The latter is required to obtain particles with radii compatible with SANS experiments (see Materials and methods for additional details on the synthesis protocol). All microgels contain a constant amount of crosslinker: 5 mol \% N,N'-methylenebis(acrylamide) (BIS).
A DLS characterization of the microgels dispersed in $H_2O$ is reported in \jv{Figure \ref{fig1}A} and in Table \ref{tbl:synthesis}. The evolution of the hydrodynamic radius ($R_h$) as a function of temperature is fitted with eq. \ref{eq:DLS} (black lines) in order to extract $T_{VPT}$ for each NIPAM-NIPMAM composition (\jv{Figure \ref{fig1}B, black circles}) and the parameter $1/s$, which represents the width of the transition (\jv{Figure \ref{fig1}C}). \fb{The full set of fitting parameters is reported in Table S1.}
Interestingly, $T_{VPT}$ of the copolymer microgels does not precisely match the expected value for a linear dependence with \jv{composition,\cite{Clarke2013,Djokpe2001} but is instead consistently approximately 1°C lower.} 
\jv{This result is further corroborated by differential scanning calorimetry (DSC) experiments (Figure S1), with $T_{VPT}$ determined from the onset of temperature increase in the thermograms.
Despite a consistent discrepancy of approximately 3 °C between $T_{VPT}$ obtained from DLS and DSC analysis, also with the latter we observe slightly lower transition temperatures for the copolymer microgels (Figure \ref{fig1}B, gray triangles).} 

\jv{Figure \ref{fig1}C additionally} shows that the transition width increases with increasing PNIPMAM content, albeit not in a linear fashion. Regarding the homopolymer PNIPAM and PNIPMAM microgels, a broader transition for the latter captures what is observed in previous works \cite{Wedel2016,cors2019deuteration}.
Overall, these results suggest that thermal responsivity depends not only on composition but also on the internal polymer distribution. However, a standard DLS \jv{and DSC} characterization does not allow for a deeper understanding of this phenomenon.


\begin{figure}[t!]
\centering
\includegraphics[scale=1]{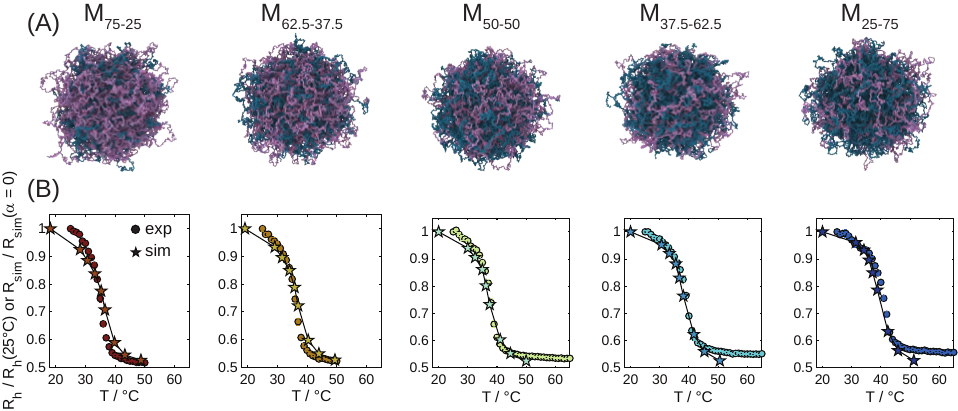}
\caption{\small \jv{Numerical simulations of PNIPAM-PNIPMAM microgels. (A) Snapshots of PNIPAM-PNIPMAM microgels with different compositions (indicated). PNIPAM beads are in pink, PNIPMAM beads in green. (B) Comparison between experimental (circles) and numerical (stars) swelling curves, plotted by normalizing, for each microgel, values of $R$ to the one in the swollen state: $R_h$(T = 25°C) for experiments for simulations or $R_{sim}$($\alpha$ = 0). The procedure to map the solvophobic parameter $\alpha$ to experimental temperature $T$ is described in the Methods.}}
\label{fig2}
\end{figure}

In parallel, we also exploited monomer-resolved simulations to obtain \textit{in-silico} copolymer microgels with varying amounts of the two monomers. Representative simulation snapshots of the microgels in their swollen state are reported in \jv{Figure \ref{fig2}A, where 
M$_{75-25}$ indicates a P(H-NIPAM-\textit{co}-H-NIPMAM) microgel with 75 mol \% NIPAM and 25 mol \% NIPMAM, and so on.} The simulations are performed as described in Methods, using the so-called ``\textit{block}'' topology, which yielded favourable results in comparison to experiments for equimolar microgels~\cite{Tavagnacco2025}.
To complement DLS \jv{and DSC}, we also monitor the swelling behavior in simulations (see Materials and methods for further details). 
\fb{The swelling curves derived from simulations are superimposed to experimental ones in Figure \ref{fig2}A. To achieve a proper comparison, for each sample we normalized the hydrodynamic radius to that of the swollen microgels. Moreover, we mapped the solvophobic parameter $\alpha$ to experimental temperature by extending the relation previously derived in ref.~\citenum{Tavagnacco2025} for microgel M$_{50-50}$ to account for the different compositions of the samples, as detailed in the Materials and methods. Noteworthy, we found that the swelling curves obtained in this way accurately reproduce the evolution of the particle size as a function of temperature and the influence of copolymer content on the swelling behavior.}

\begin{figure}[t!]
\centering
\includegraphics[scale=0.5]{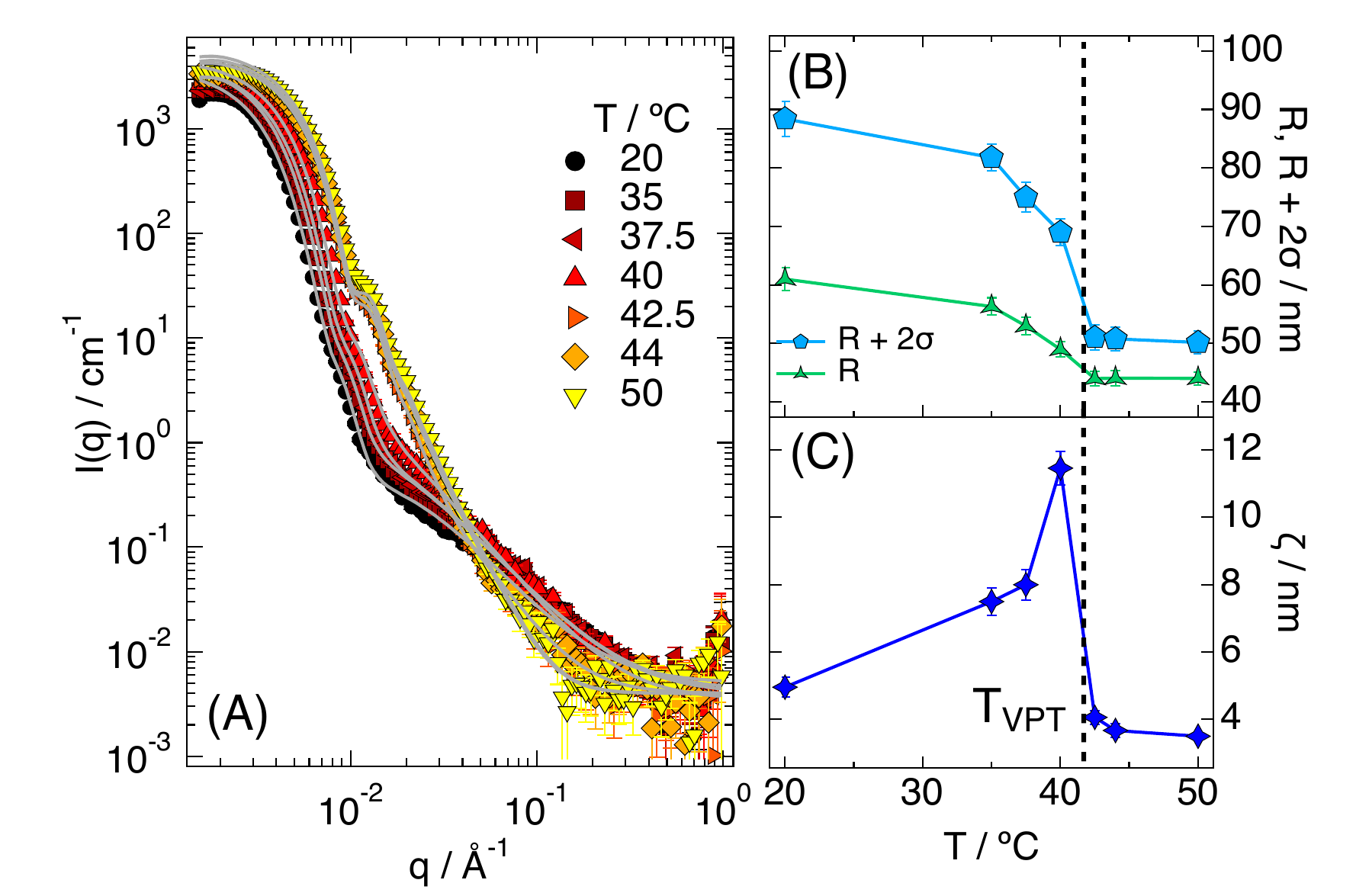}
\caption{\small SANS of M$_{25-75}$. A) SANS scattering intensities $I(q)$ as a function of temperature. Lines represent fits with the fuzzy sphere model in eq. \ref{eq:Pfuzzy}. B) Particle radii $R$ and $R + 2\sigma$ from SANS fits as a function of temperature. C) Correlation length of the polymer mesh ($\zeta$). In B) and C) \jv{the dashed line indicates $T_{VPT}$} from DLS analysis.}
\label{fig3}
\end{figure}

For an in-depth characterization of the internal polymer network conformation as a function of NIPMAM comonomer ratio, we performed SANS measurements of dilute suspensions in \ce{D2O} to determine the particle form factors. In \jv{Figure \ref{fig3}} we report the scattering intensities $I(q)$ of the exemplary M$_{25-75}$ for temperatures spanning its VPT. SANS data for the other PNIPMAM compositions are in the Supporting Information (\jv{Figures S2-S14}).
The well-known fuzzy sphere model from Stieger et al. \cite{Stieger2004} (eq. \ref{eq:Pfuzzy}) nicely captures the experimental form factors for all compositions and investigated temperatures. \jv{We note that in our experiments we do not discern non-thermoresponsive domains in the homopolymer cases, as recently visualized by using high-speed atomic force microscopy on bigger particles.\cite{Matsui2018,Nishizawa2019}}
As shown in \jv{Figure \ref{fig3}A}, we observe the expected decrease in size at high temperature, indicated by a right-shift of the Guinier \jv{(low-q)} region of the scattering curves. This is coupled with an increase in the slope of $I(q)$ at higher $q$ values, \jv{indicative of a transition from a more open polymer network with a diffuse interface to one that is collapsed and with a more compact interface.}

The resulting particle radius from the form factor fits is plotted in \jv{Figure \ref{fig3}B} and shows particle deswelling up to $T_{VPT}$, after which the microgel is collapsed and the contribution of the particle fuzziness $2\sigma$ to the total radius is significantly decreased.

While this result is in line with the form factor of commonly investigated PNIPAM microgels, \cite{Stieger2004} insight into the internal structure can be obtained by looking at the variation in the correlation length of the polymer mesh ($\zeta$) as a function of temperature (\jv{Figure \ref{fig3}C}). Interestingly, $\zeta$ first increases, reaching a maximum at $T = 40^\circ C$, and then sharply decreases above $T_{VPT}$, which is 40.9 $\pm$ 0.1 °C for this microgel, when the network is significantly collapsed.
We interpret $\zeta$ as an estimate of the average distance between heterogeneous, non collapsed regions within the network. $\zeta$ increases near $T_{VPT}$ due to the onset of network collapse, which increases the overall heterogeneity of the microgel architecture, comprised of collapsed and still swollen domains.
This trend of $\zeta$ increasing below $T_{VPT}$ and then sharply decreasing afterwards is consistently reproduced for all investigated microgels (\jv{Figures S2-S14}).

In order to challenge the peculiar block structure made of separate PNIPAM and PNIPMAM domains that we put forward for describing the internal architecture of 50-50 P(NIPAM-\textit{co}-NIPMAM) microgels, \cite{Tavagnacco2025} in \jv{Figures S15 and S16} we compare experimental form factors for M$_{25-75}$ and for a 25-75 P(D-NIPAM-\textit{co}-H-NIPMAM) microgel, 
with numerical ones obtained with the \textit{random} or \textit{block} models (see Materials and methods for details on the comparison procedure). 
As previously discussed, \cite{Tavagnacco2025} isotopic labeling of one of the two polymers for contrast matching in $D_2O$ is mandatory in order to distinguish between \textit{random} or \textit{block} topologies and gain insights on the internal structure. Indeed, \jv{Figure S15} shows that both the \textit{random} and \textit{block} architectures capture the evolution of the form factors across $T_{VPT}$ for the fully hydrogenated M$_{25-75}$ sample, not being able to provide details on the internal architecture. Instead, \jv{Figure S16} highlights an excellent agreement of the numerical data in describing the microgel architecture when the \textit{block} model is used, while the \textit{random} model provides a less accurate representation of experimental data, and therefore validating the \textit{block} topology also for P(NIPAM-\textit{co}-NIPMAM) microgels of different composition.
Additionally, the resulting $\zeta$ values are much more moderate for the D-H microgel with respect to the H-H one (\jv{Figure S17}), hinting at a more homogeneous network. Since in D-H microgels in D$_2$O the D-PNIPAM domains are contrast-matched with the solvent and do not contribute to the scattering profile, this indicates that the PNIPMAM network is more homogeneous than the PNIPAM one, and corroborates the attribution of network heterogeneities observed in H-H microgels to the formation of PNIPAM domains at the deswelling transition.

\bigskip

\begin{figure}[t!]
\centering
\includegraphics[scale=0.55]{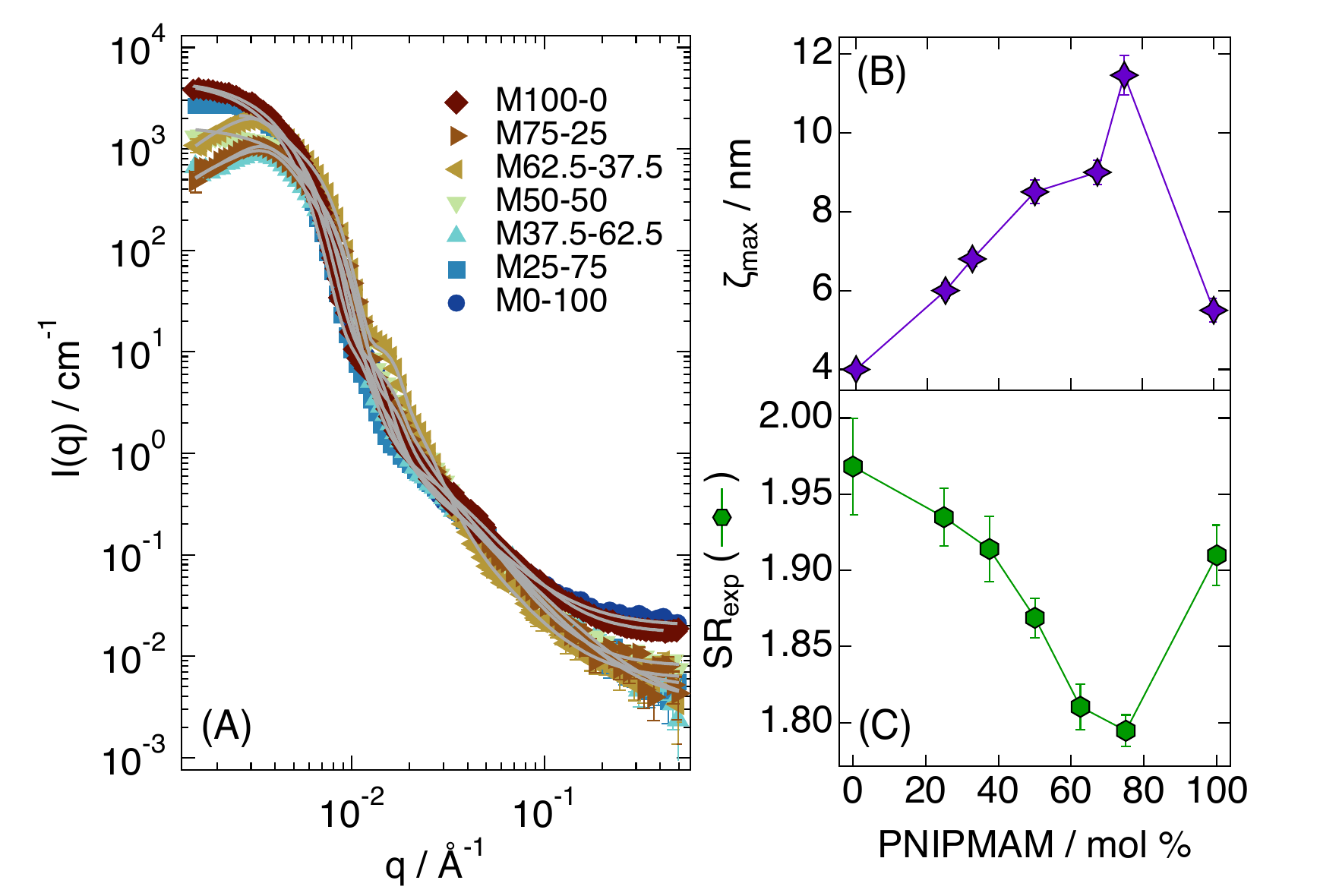}
\caption{\small A) SANS scattering intensities $I(q)$ for different compositions at the temperature at which the correlation length of the polymer mesh is maximum. Lines are fits with the fuzzy sphere model. B) Maximum correlation length $\zeta_{max}$ and (C) swelling ratio from experiments ($SR_{exp}$) as a function of PNIPMAM mole fraction.}
\label{fig4}
\end{figure}

We then investigated in detail how the NIPMAM mole fraction affects the internal polymer distribution and reorganization as a function of temperature. In \jv{Figure \ref{fig4}A} we compare SANS scattering intensities measured at the temperature corresponding to the maximum in the correlation length of the polymer mesh ($\zeta_{max}$), for all investigated compositions. We note that such temperature increases with increasing NIPMAM content.
As already indicated, we interpret $\zeta_{max}$ as an estimate of the average distance between non collapsed regions within the network, \jv{thereby} giving information on the internal heterogeneity. The resulting $\zeta_{max}$ (\jv{Figure \ref{fig4}B}) is the lowest for pure PNIPAM or PNIPMAM microgels, indicating that the homopolymer networks have an overall more homogeneous polymeric distribution at the VPT, as expected for particles composed of a single monomer. The discrepancy that is observed between pure PNIPAM and pure PNIPMAM microgels hints to a slight increase in network heterogeneity during collapse for the latter monomer, in line with the increased transition width (\jv{Figure \ref{fig1}C}).
In copolymer networks, $\zeta_{max}$, by providing information on the network heterogeneity, is therefore an estimate of the average distance between NIPAM-rich domains, which collapse first due to their lower critical solution temperature with respect to PNIPMAM. Upon increasing the NIPMAM mole fraction, the average distance between collapsed NIPAM-rich blocks progressively increases. Below 50 \% there is still a significant amount of NIPAM in the network, resulting in a few non-collapsed regions and an intermediate value of $\zeta_{max}$. Instead, a further decrease in NIPAM content yields networks with only a few collapsed regions, characterized by a further increase in $\zeta_{max}$, which reaches its maximum for M$_{25-75}$ among the studied range of compositions.

In \jv{Figure \ref{fig4}C} we display the microgel's \fb{experimental} swelling ratio (\jv{$SR_{exp}$, green points}) measured as the ratio between the hydrodynamic radii $R_h (25^\circ C) / R_h (T_{max})$, where $T_{max}$ is chosen to be well-above $T_{VPT}$ in order to capture fully collapsed particles: 50°C up to 37.5 \% NIPMAM mol fraction, 65°C for the remaining particles. Interestingly, $SR_{exp}$ as a function of NIPMAM content mirrors the $\zeta_{max}$ trend, with a continuous decrease up to a minimum for M$_{25-75}$, which corresponds to the more heterogeneous structure prior to network collapse, and a subsequent increase for M$_{0-100}$. 
The swelling ratio is a thermodynamic quantity measured at equilibrium between fully swollen and fully collapsed particles. This is therefore a surprising result, indicating that the collapsed microgels maintain a memory of the transition, with a decreased deswelling for more heterogeneous networks even well-above their respective VPT. 
The smaller deswelling indicates that copolymer microgels detain more water at high temperatures. As a consequence, they are effectively stiffer with respect to their temperature response, being more able to resist the induced collapse.
We also note that once again the difference between homopolymer PNIPAM and PNIPMAM microgels is in line with what observed in the literature. \cite{Wedel2016,cors2019deuteration}

\section{Conclusions}
We have combined DLS, \jv{DSC,} SANS and monomer-resolved numerical simulations to unravel how comonomer ratio governs the internal architecture and resulting thermoresponsive behavior of P(NIPAM-\textit{co}-NIPMAM) copolymer microgels. While $T_{VPT}$ increases with increasing NIPMAM content, its deviation from the expected value in case of a linear dependence on composition, as well as the non-monotonic broadening of the transition, hint that the polymer network conformation changes in a non-trivial way with composition and in response to the stimulus. 
Insights at the nanoscale were obtained from SANS analysis. The temperature evolution of the polymer mesh correlation length $\zeta$, obtained from fitting of the form factors, indicates an increase in network heterogeneity at the onset of collapse, followed by a sharp decrease in the collapsed state. The increase of $\zeta$ below $T_{VPT}$ is universal for all compositions investigated and identifies a regime of enhanced structural heterogeneity associated with the coexistence of collapsed and swollen domains.

Exploiting isotopic labelling and comparison with simulations, we provide evidence that radical copolymerization leads to the formation of a block-like internal topology composed of small domains containing mainly NIPAM or NIPMAM, rather than a random monomer distribution, as already pointed out in the case of P(NIPAM-\textit{co}-NIPMAM) microgels of equimolar compositions. \cite{Tavagnacco2025} 
As a consequence, $\zeta$ can be interpreted as the average spacing between NIPAM-rich domains that collapse first owing to their lower VPTT. The non-monotonic dependence of $\zeta_{max}$ with composition, peaking at 75 mol \% NIPMAM, demonstrates that copolymer microgels develop maximal internal segregation when NIPAM is present in a minor yet sufficient amount to form discrete responsive domains, which are separated by increasingly large non-collapsing PNIPMAM regions.  
Interestingly, the $\zeta_{max}$ trend is mirrored by the equilibrium swelling ratio, both experimentally and in silico, revealing that the fully collapsed state retains a structural memory of the heterogeneous transition pathway. Specifically, copolymer microgels with increased heterogeneity exhibit reduced deswelling, indicating an effectively less compliant network in response to temperature variations. Future works could focus on quantifying the microgel's bulk modulus as a function of NIPMAM content investigating whether the decreased swelling ratio for more heterogeneous networks translates into stiffer particles also in their swollen state. 

Overall, these findings establish a direct link between mesoscale internal organization and macroscopic thermodynamic response, highlighting that copolymer microgel functionality is encoded not only in its chemical composition but also in the spatial monomer distribution, a detail often overlooked in copolymer systems. More broadly, this indicates how engineering the internal architecture allows tailoring their response by modulating their transition and deswelling, thereby enabling fine control over mechanical properties and stimuli responsiveness of interest in applications ranging from sensing to drug delivery and additive manufacturing.

\begin{acknowledgement}

The authors thank M. Bertoldo for valuable discussions.
We acknowledge RAL for time at the SANS2D beamline under proposals 2310267 and 2320109 (data available at DOI: https://doi.org/10.5286/ISIS.E.RB2310267 and https://doi.org/10.5286/ISIS.E.RB2320109). We acknowledge Najet Mahmoudi for assistance during the experiments.
This work is based on experiments performed at the Swiss spallation neutron source SINQ, Paul Scherrer Institute, Villigen, Switzerland, instrument SANS-I, proposal 20220847. We acknowledge Urs Gasser for assistance during the experiments.
\jv{This work benefited from the use of the SasView project (www.sasview.org), originally developed under NSF award DMR-0520547 and containing code developed with funding from the European Union’s Horizon 2020 research and innovation programme under the SINE2020 project, grant agreement No 654000.}
The authors gratefully acknowledge the financial support of Consiglio Nazionale delle Ricerche within CNR-STFC Agreement 2021-2027 (N 0065606), concerning collaboration in scientific research 2310267 and 2320109 at the ISIS Neutron and Muon Source (UK) of Science and Technology Facilities Council (STFC).
J.V. acknowledges funding from Ministero dell'Università e della Ricerca through the Rita Levi Montalcini research program (D.M. 1317 published on 15-12-2021, grant Nr. PGR21W3GY8).
M.L. and E.Z. acknowledge financial support by Progetto Co-MGELS funded by the European Union - NextGeneration EU under the National Recovery and Resilience Plan Mission 4 “Istruzione e Ricerca” - Component C2 - Investment 1.1 - “Fondo PRIN”, Project code PRIN2022PAYLXW Sector PE11, CUP B53D23008890006.

\end{acknowledgement}

\section*{Conflict of Interest}
The authors declare no conflict of interest.

\begin{suppinfo}

\jv{The Supporting Information is available free of charge at:
\\
The Supporting Information contains: values of the parameters obtained by fitting the experimental swelling from DLS data; DSC thermograms; SANS scattering intensities, fits and resulting parameters as a function of temperature for all the investigated copolymer microgels; comparisons between experimental and numerical form factors; correlation length of the polymer mesh ($\zeta$) for M$_{25-75}$ microgels containing H-NIPAM or D-NIPAM.}


\end{suppinfo}

\bibliography{main_revised_arxiv}

\newpage

\begin{Large}
\section{Supporting information for:}
\end{Large}

\bigskip

\begin{center}
\begin{LARGE}
\textbf{Heterogeneous collapse in thermoresponsive copolymer microgels varying molar composition}
\end{LARGE}
\end{center}

\setcounter{figure}{0}
\renewcommand{\thefigure}{S\arabic{figure}}
\renewcommand{\thetable}{S\arabic{table}}

\newpage

\begin{table}[!htb]
    \centering
    \small
    \begin{tabular}{lcccccc}
        \hline
        sample & $T_\text{VPT}$ & $1/s$ & $R_{max}$ & $\Delta R$ & $A$     & $B$                     \\
                 & (°C)           & (°C)  & (nm)      & (nm)       & (nm/°C) & ($10^{-3}$ nm/°C$^2$) \\
        \hline
        M$_{100-0}$     & 33.1 $\pm$ 0.1 & 2.36 $\pm$ 0.13 & 41.4 $\pm$ 0.1 & \;\,9.3 $\pm$ 0.3 & -0.45 $\pm$ 0.02 &    13.9 $\pm$ 1.0 \\  
        M$_{75-25}$     & 35.0 $\pm$ 0.1 & 2.73 $\pm$ 0.12 & 53.2 $\pm$ 0.2 &    13.6 $\pm$ 0.4 & -0.32 $\pm$ 0.03 &    10.1 $\pm$ 1.3 \\    
        M$_{62.5-37.5}$ & 36.1 $\pm$ 0.1 & 2.67 $\pm$ 0.13 & 59.2 $\pm$ 0.2 &    14.4 $\pm$ 0.4 & -0.36 $\pm$ 0.04 &    11.4 $\pm$ 1.5 \\    
        M$_{50-50}$     & 37.9 $\pm$ 0.1 & 2.60 $\pm$ 0.11 & 63.3 $\pm$ 0.1 &    14.5 $\pm$ 0.3 & -0.36 $\pm$ 0.02 & \;\,8.6 $\pm$ 0.6 \\    
        M$_{37.5-62.5}$ & 39.2 $\pm$ 0.1 & 2.68 $\pm$ 0.11 & 52.7 $\pm$ 0.1 &    11.1 $\pm$ 0.2 & -0.31 $\pm$ 0.01 & \;\,7.5 $\pm$ 0.4 \\    
        M$_{25-75}$     & 40.9 $\pm$ 0.1 & 2.98 $\pm$ 0.12 & 75.9 $\pm$ 0.2 &    16.7 $\pm$ 0.3 & -0.30 $\pm$ 0.03 & \;\,7.2 $\pm$ 0.8 \\    
        M$_{0-100}$     & 44.4 $\pm$ 0.1 & 3.90 $\pm$ 0.18 & 69.6 $\pm$ 0.2 &    18.2 $\pm$ 0.5 & -0.22 $\pm$ 0.03 & \;\,8.1 $\pm$ 0.8 \\    
        \hline
    \end{tabular}
    \caption{Fitting parameters of the experimental swelling curve to eq.~1.}
    \label{tab:fit}
\end{table}
\noindent Table~\ref{tab:fit} reports the values of the parameters obtained by fitting the experimental swelling curves of microgels at varying co-polymer compositions. The data and the fitting curves are plotted in Fig.~1B of the main text. The parameters $T_\text{VPT}$ and $1/s$ are plotted as a function of the molar fraction of NIPMAM in Figs.~1C and 1D, respectively. The coefficients A and B, weight the linear and quadratic phenomenological terms in eq.~1, allowing the fitting function to accurately reproduce the experimental data while preserving the physical meaning of the relevant fitting parameters.

\newpage


\begin{figure}[!htb]
\centering
\includegraphics[scale=0.6]{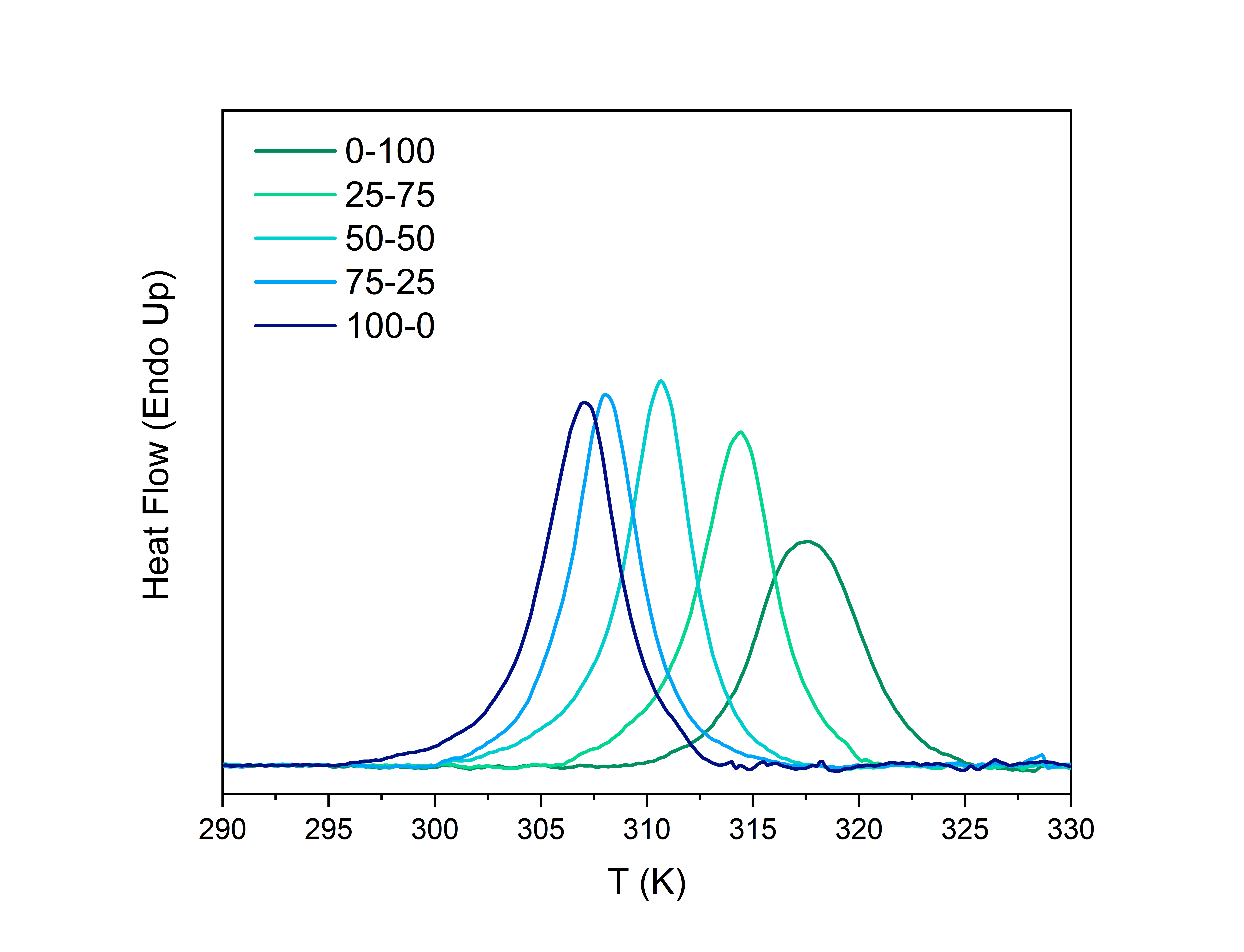}
\caption{\small \textbf{Differential scanning calorimetry.} DSC thermograms for P(NIPAM-\textit{co}-NIPMAM) microgels at varying mol fraction of the components, as indicated.}
\label{fig:M0}
\end{figure}

\begin{figure}[!htb]
\centering
\includegraphics[scale=0.6]{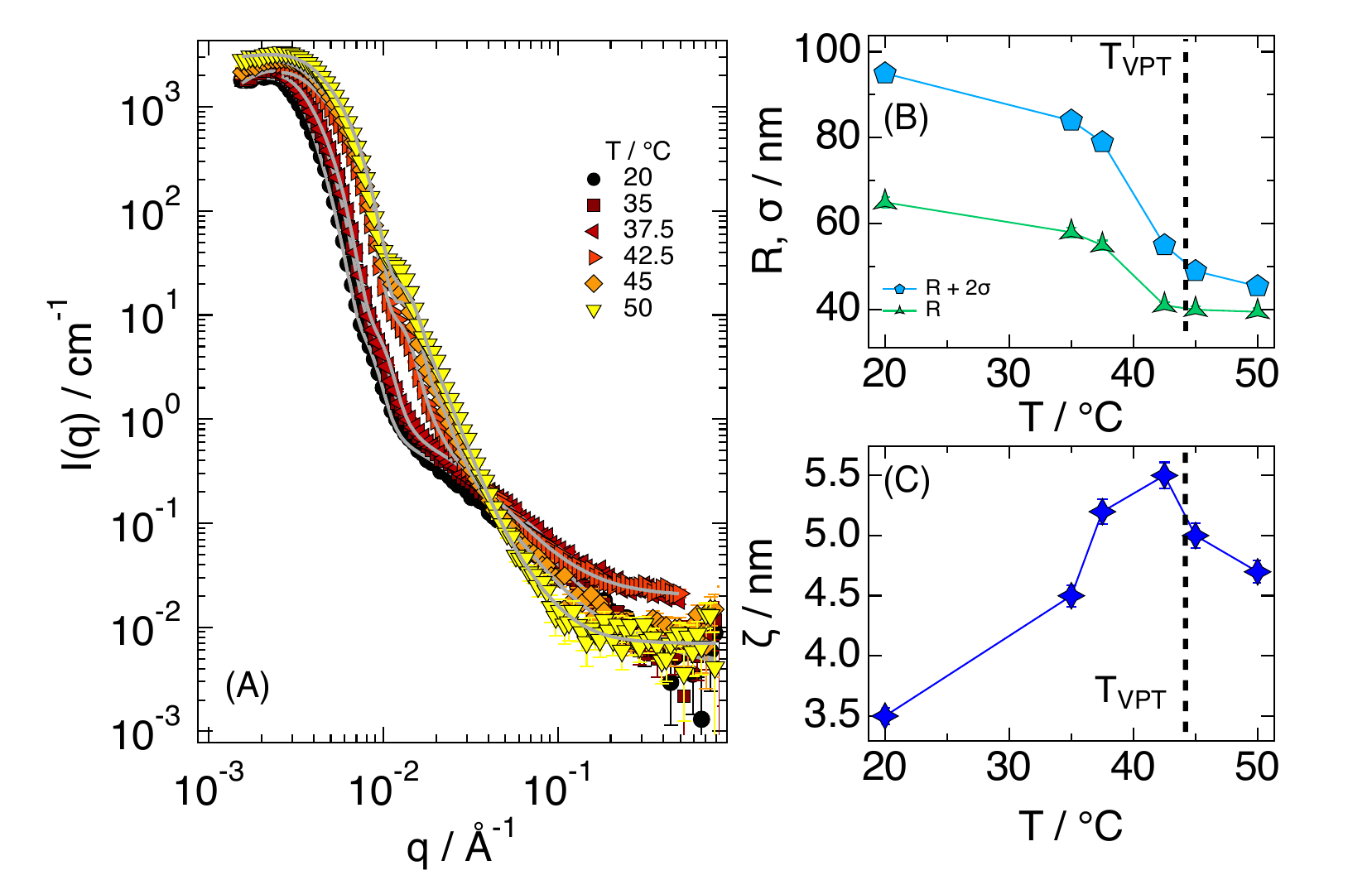}
\caption{\small \textbf{SANS of M$_{0-100}$.} Left: SANS scattering intensities $I(q)$ as a function of temperature. Lines represent fits with the fuzzy sphere model, including $S(q)$ contribution. Top-right: particle core radius ($R$) and total radius ($R + 2\sigma$) from SANS fits as a function of temperature. Bottom-right: correlation length of the polymer mesh ($\zeta$). $T_{VPT}$ (dashed line) indicates the VPTT from DLS analysis.}
\label{fig:M0}
\end{figure}

\begin{figure}[!htb]
\centering
\includegraphics[scale=0.4]{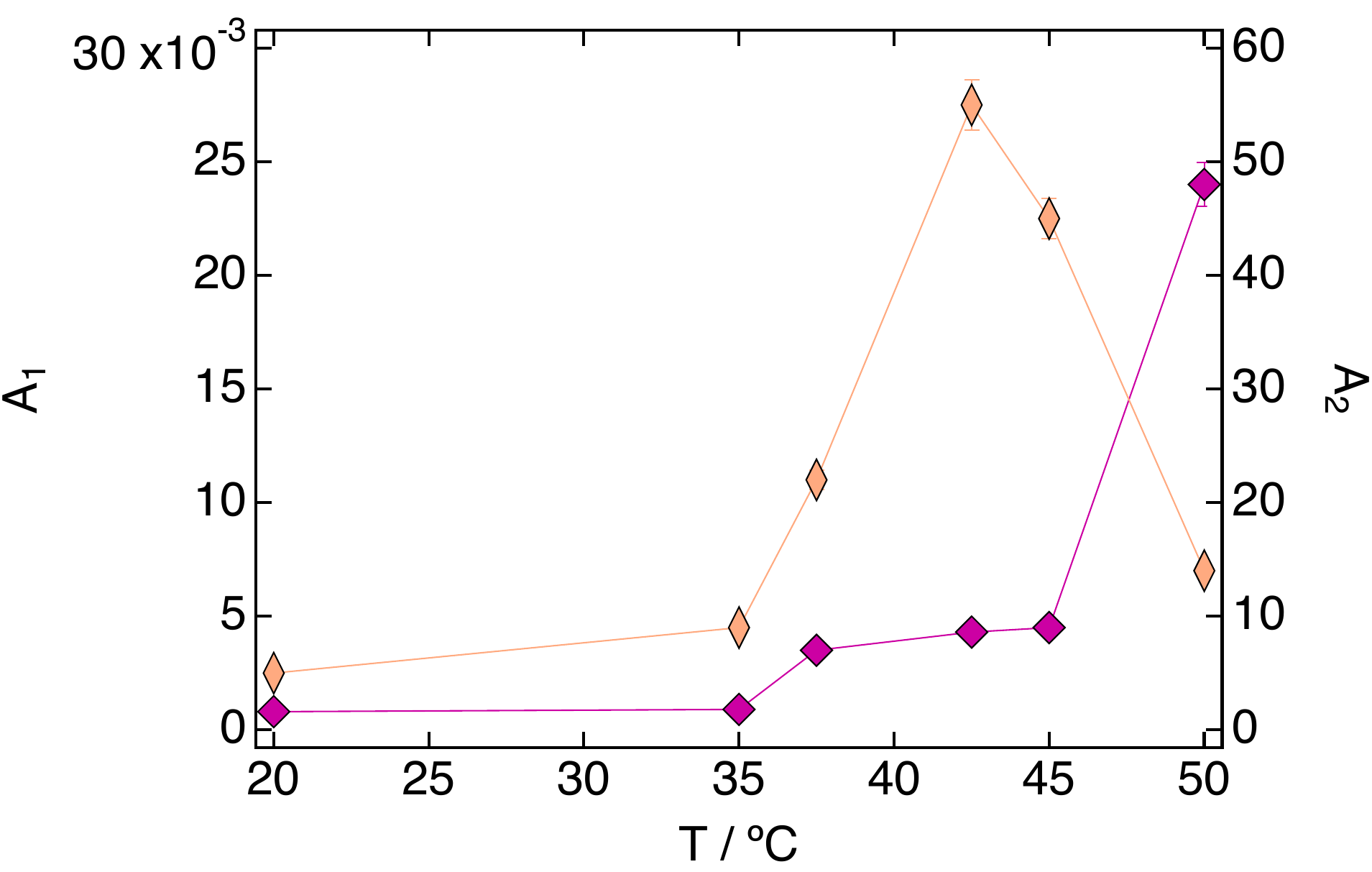}
\caption{\small \textbf{SANS of M$_{0-100}$.} Parameters $A_1$ (pink squares) and $A_2$ (orange diamonds) from SANS fits as a function of temperature.}
\label{fig:M0_A1A2}
\end{figure}

\begin{figure}[!htb]
\centering
\includegraphics[scale=0.4]{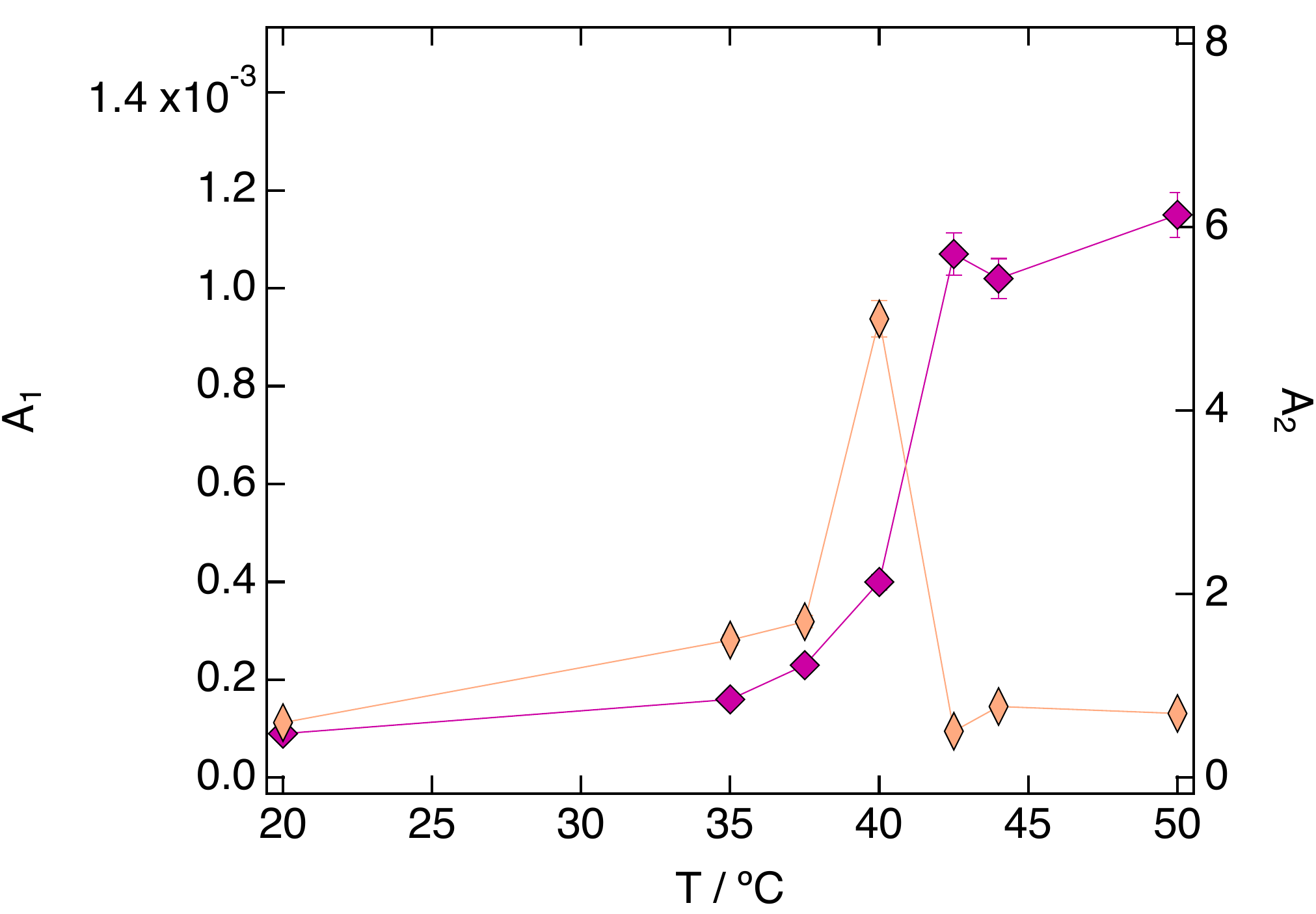}
\caption{\small \textbf{SANS of M$_{25-75}$.} Parameters $A_1$ (pink squares) and $A_2$ (orange diamonds) from SANS fits as a function of temperature.}
\label{fig:M25_A1A2}
\end{figure}

\begin{figure}[!htb]
\centering
\includegraphics[scale=0.6]{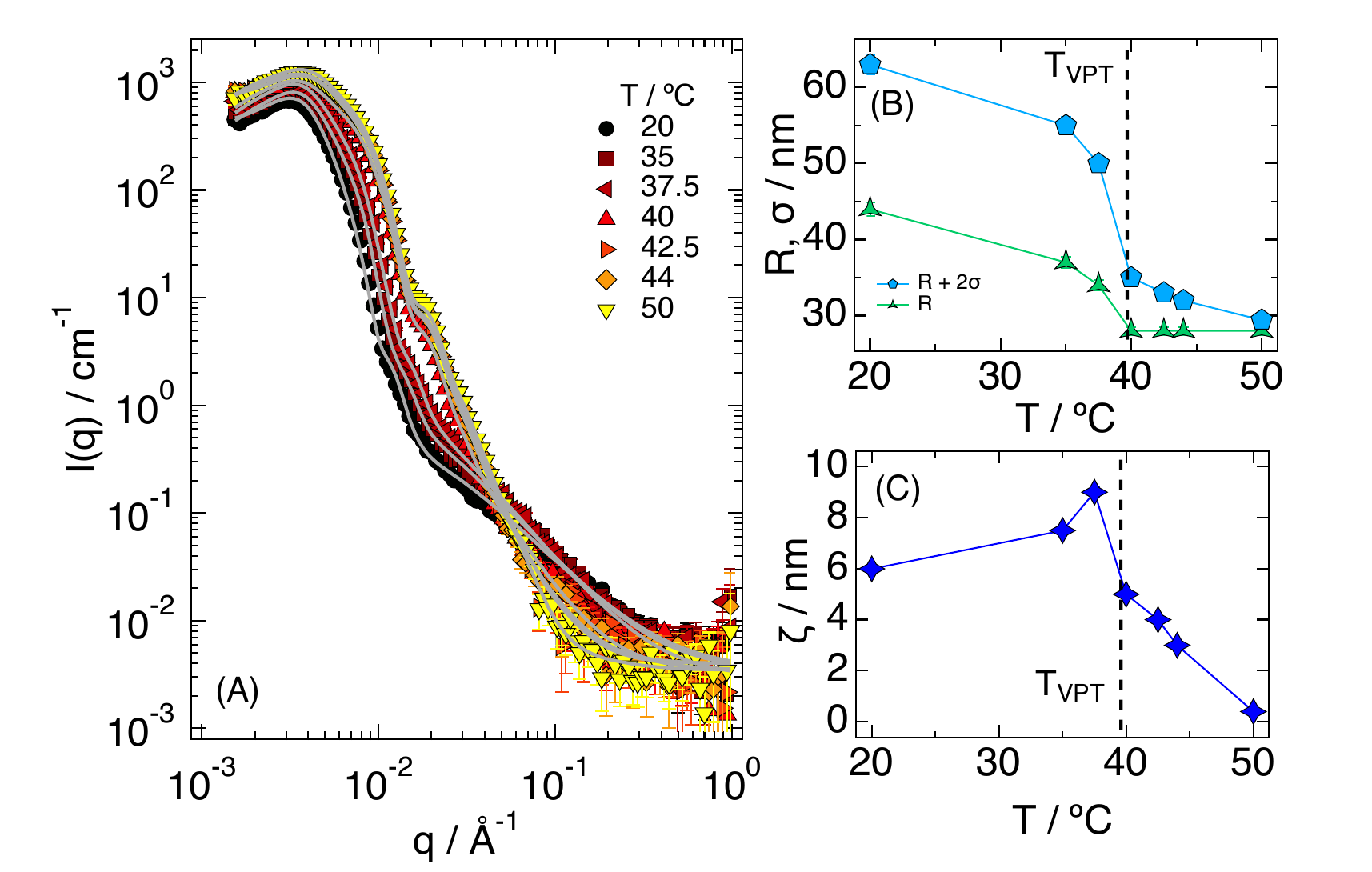}
\caption{\small \textbf{SANS of M$_{37.5-62.5}$.} Left: SANS scattering intensities $I(q)$ as a function of temperature. Lines represent fits with the fuzzy sphere model, including $S(q)$ contribution. Top-right: particle core radius ($R$) and total radius ($R + 2\sigma$) from SANS fits as a function of temperature. Bottom-right: correlation length of the polymer mesh ($\zeta$). $T_{VPT}$ (dashed line) indicates the VPTT from DLS analysis.}
\label{fig:M37p5}
\end{figure}

\begin{figure}[!htb]
\centering
\includegraphics[scale=0.4]{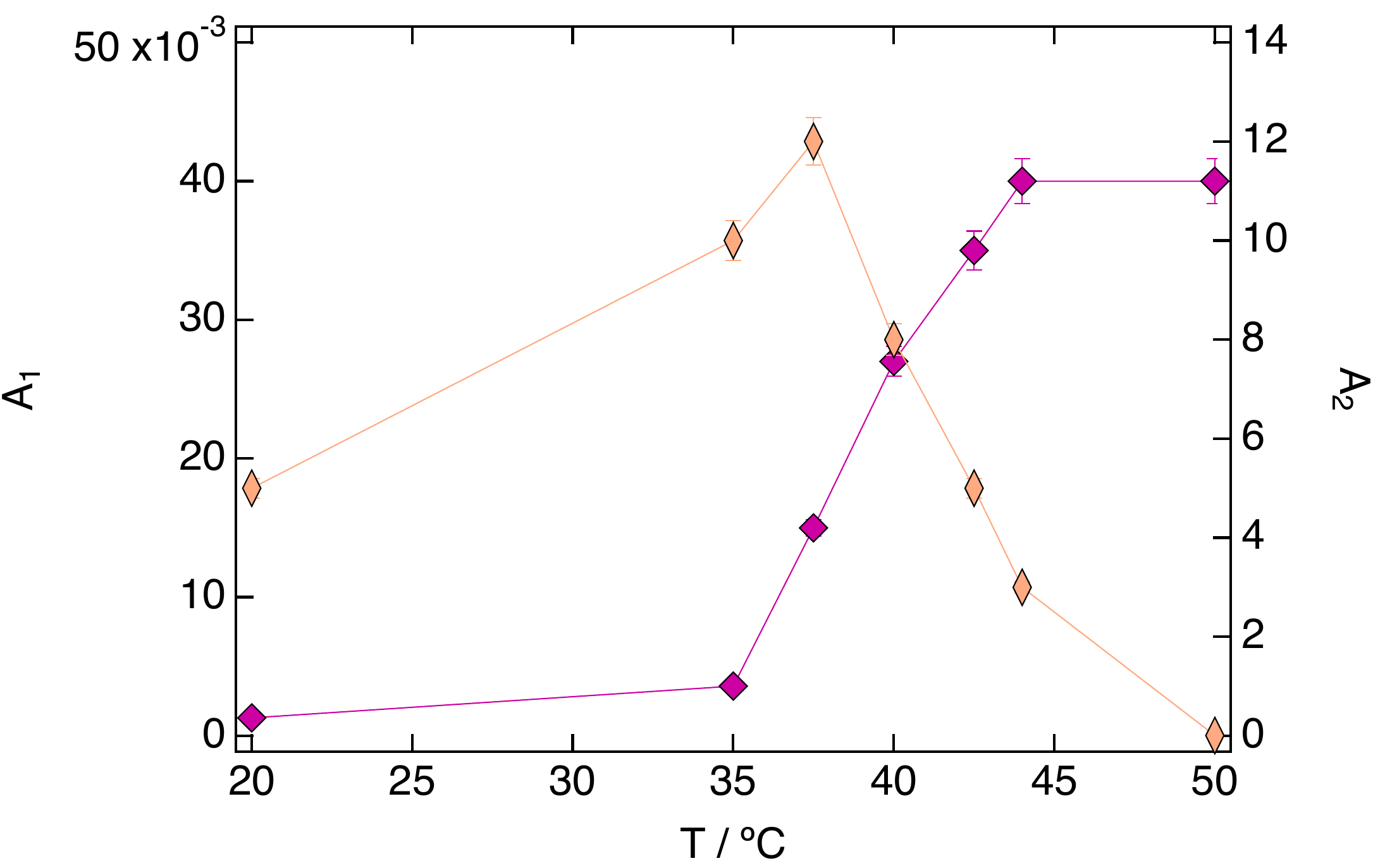}
\caption{\small \textbf{SANS of M$_{37.5-62.5}$.} Parameters $A_1$ (pink squares) and $A_2$ (orange diamonds) from SANS fits as a function of temperature.}
\label{fig:M37p5_A1A2}
\end{figure}

\begin{figure}[!htb]
\centering
\includegraphics[scale=0.6]{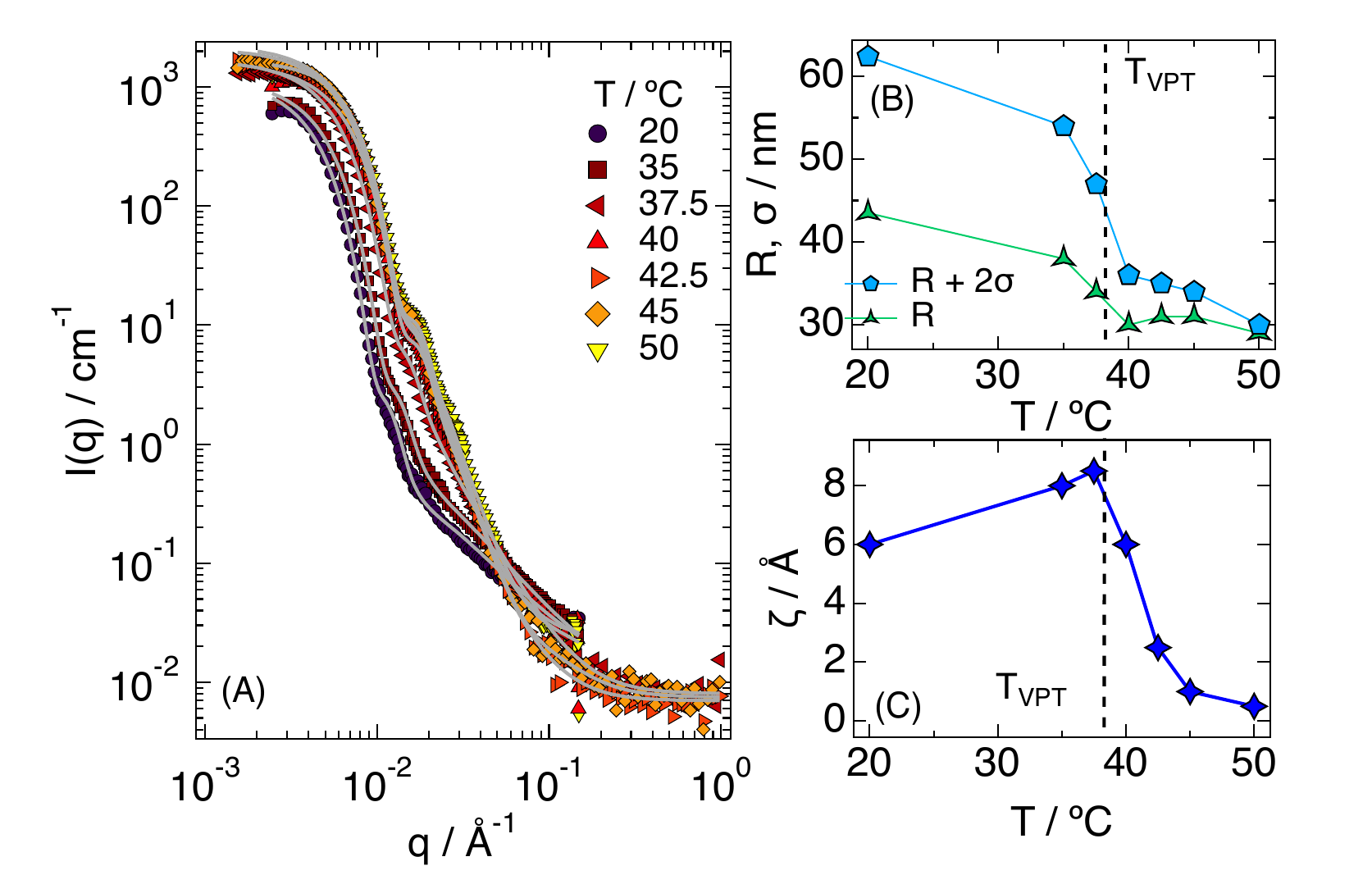}
\caption{\small \textbf{SANS of M$_{50-50}$.} Left: SANS scattering intensities $I(q)$ as a function of temperature. Lines represent fits with the fuzzy sphere model. Top-right: particle core radius ($R$) and total radius ($R + 2\sigma$) from SANS fits as a function of temperature. Bottom-right: correlation length of the polymer mesh ($\zeta$). $T_{VPT}$ (dashed line) indicates the VPTT from DLS analysis.}
\label{fig:M50}
\end{figure}

\begin{figure}[!htb]
\centering
\includegraphics[scale=0.4]{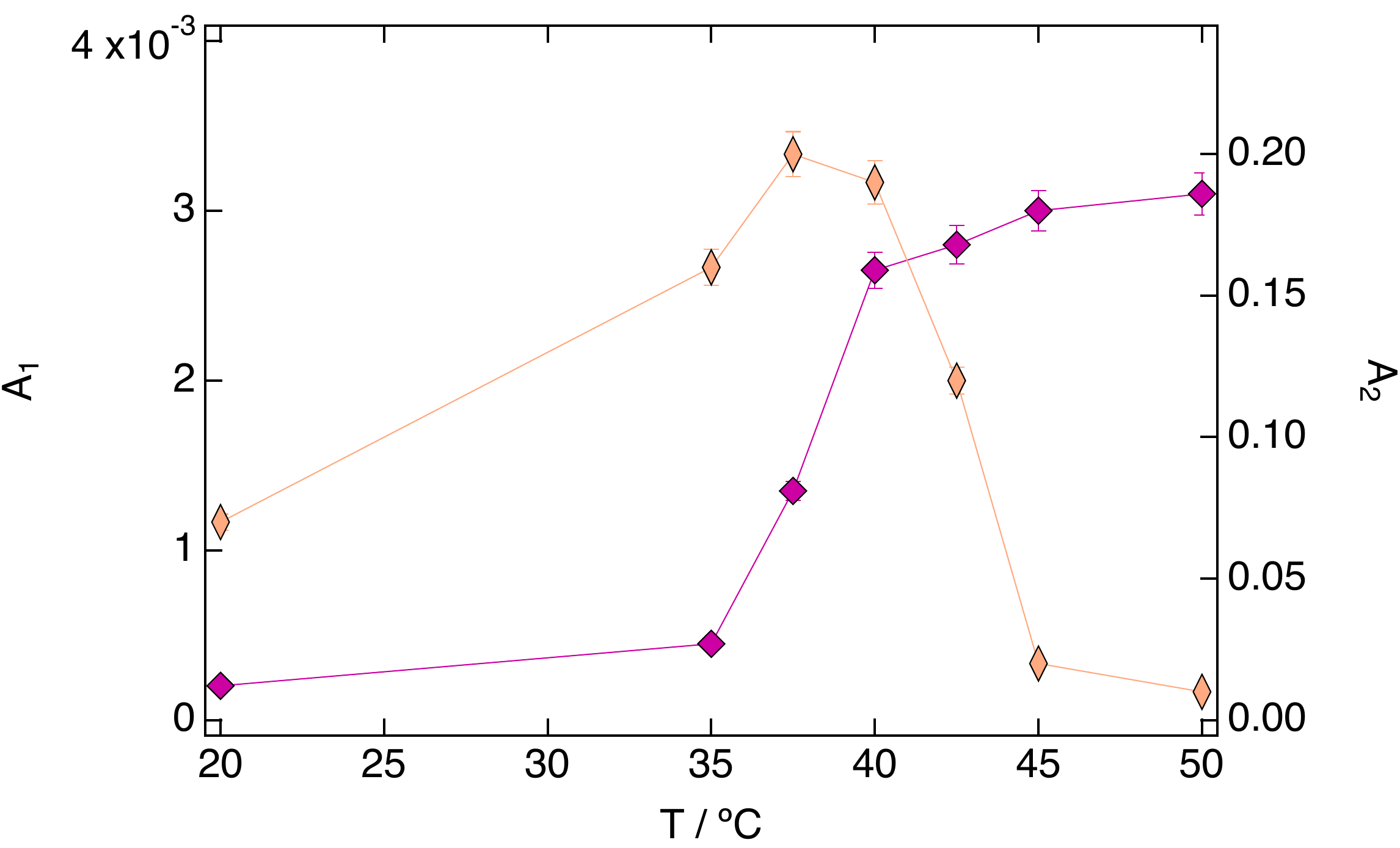}
\caption{\small \textbf{SANS of M$_{50-50}$.} Parameters $A_1$ (pink squares) and $A_2$ (orange diamonds) from SANS fits as a function of temperature.}
\label{fig:M50_A1A2}
\end{figure}

\begin{figure}[!htb]
\centering
\includegraphics[scale=0.6]{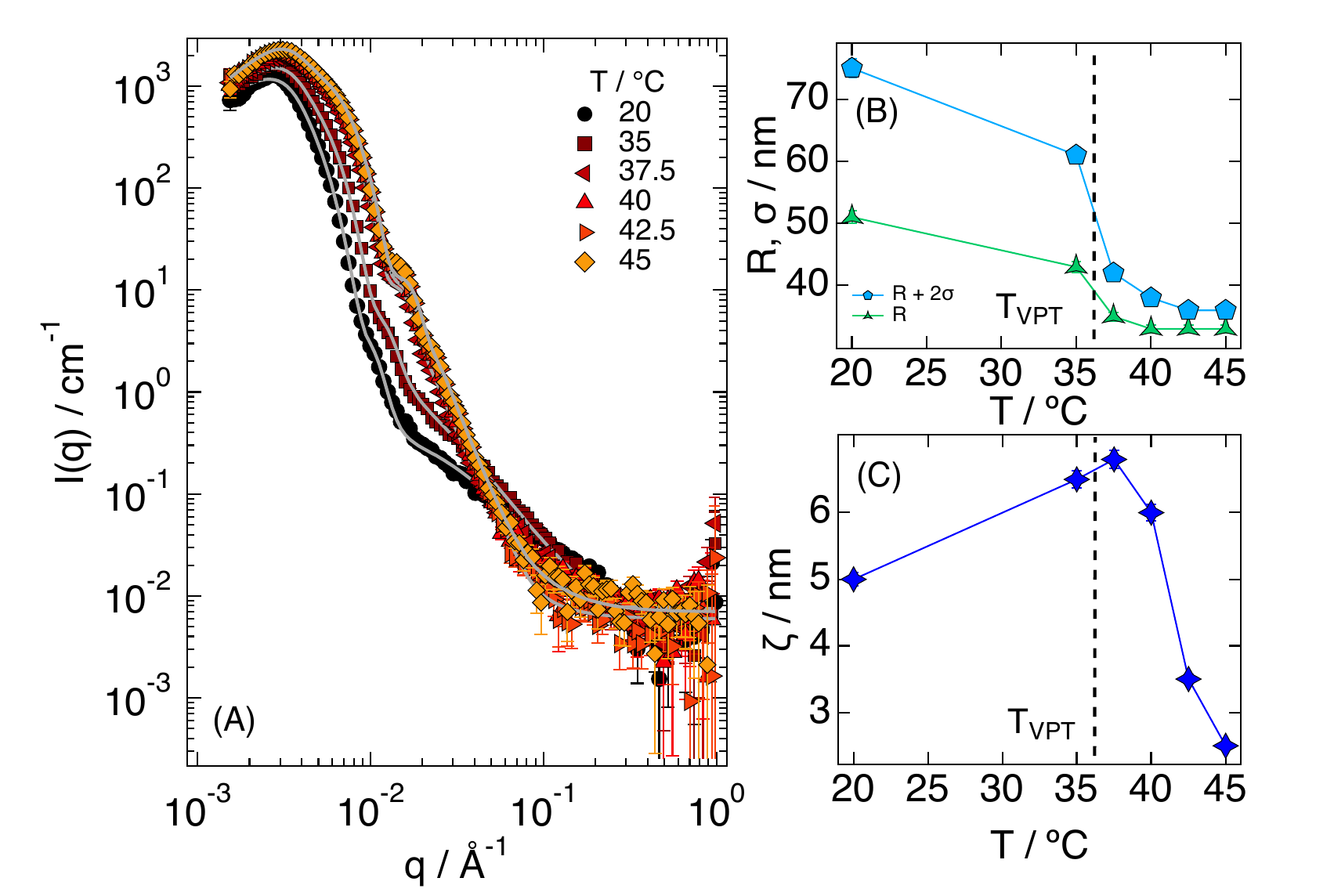}
\caption{\small \textbf{SANS of M$_{62.5-37.5}$.} Left: SANS scattering intensities $I(q)$ as a function of temperature. Lines represent fits with the fuzzy sphere model, including $S(q)$ contribution. Top-right: particle core radius ($R$) and total radius ($R + 2\sigma$) from SANS fits as a function of temperature. Bottom-right: correlation length of the polymer mesh ($\zeta$). $T_{VPT}$ (dashed line) indicates the VPTT from DLS analysis.}
\label{fig:M62p5}
\end{figure}

\begin{figure}[!htb]
\centering
\includegraphics[scale=0.4]{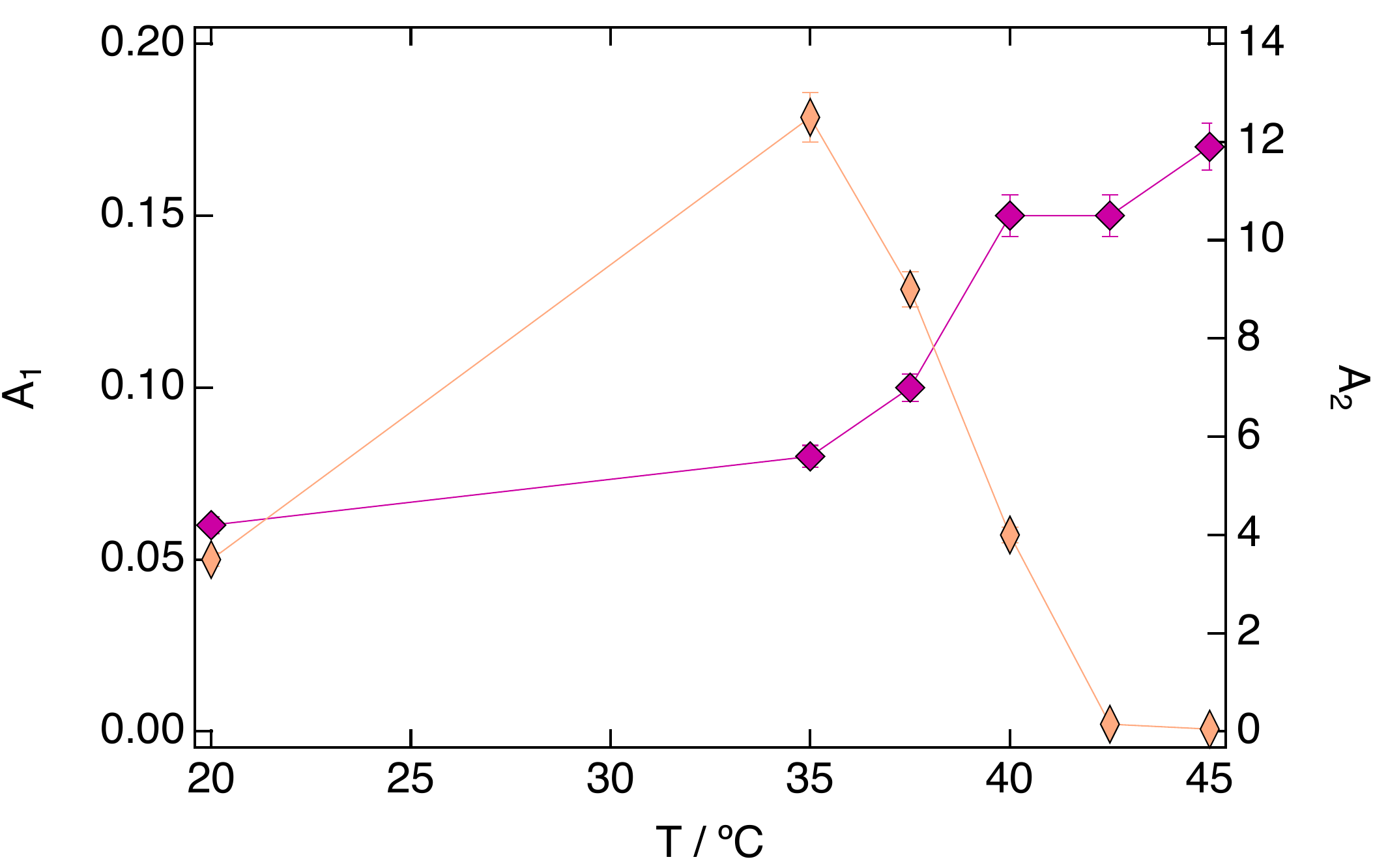}
\caption{\small \textbf{SANS of M$_{62.5-37.5}$.} Parameters $A_1$ (pink squares) and $A_2$ (orange diamonds) from SANS fits as a function of temperature.}
\label{fig:M62p5_A1A2}
\end{figure}

\begin{figure}[!htb]
\centering
\includegraphics[scale=0.6]{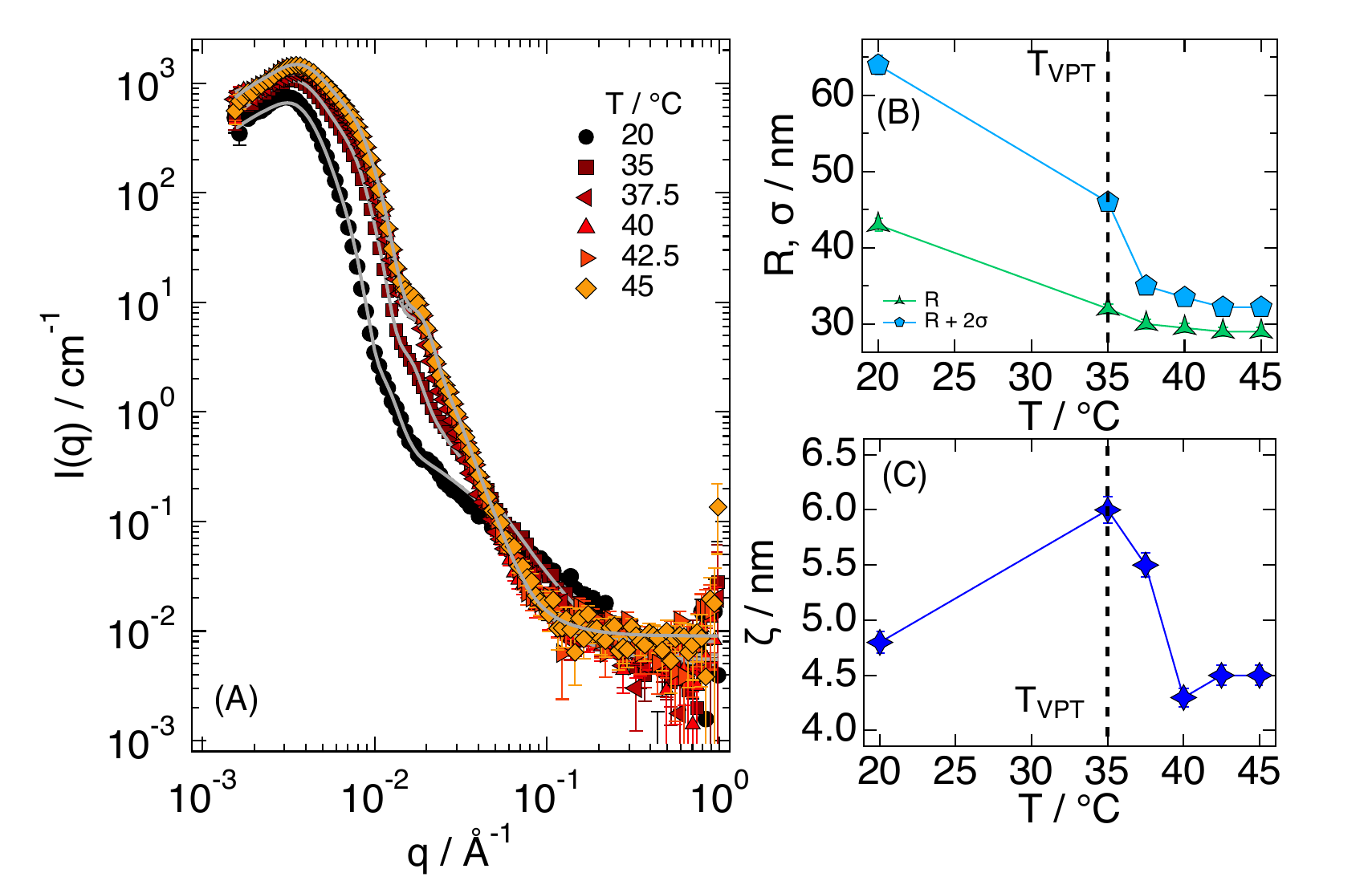}
\caption{\small \textbf{SANS of M$_{75-25}$.} Left: SANS scattering intensities $I(q)$ as a function of temperature. Lines represent fits with the fuzzy sphere model, including $S(q)$ contribution. Top-right: particle core radius ($R$) and total radius ($R + 2\sigma$) from SANS fits as a function of temperature. Bottom-right: correlation length of the polymer mesh ($\zeta$). $T_{VPT}$ (dashed line) indicates the VPTT from DLS analysis.}
\label{fig:M75}
\end{figure}

\begin{figure}[!htb]
\centering
\includegraphics[scale=0.4]{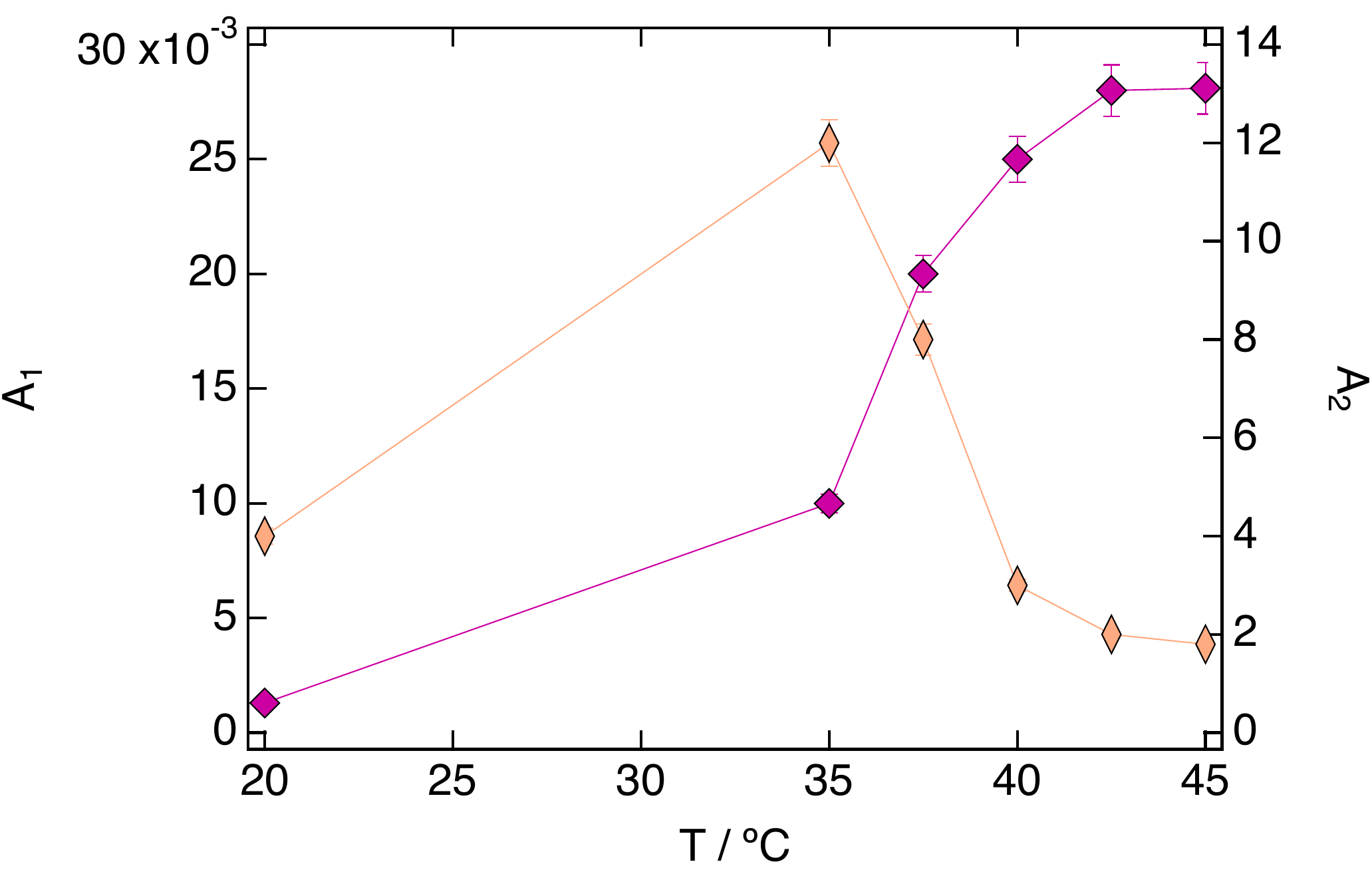}
\caption{\small \textbf{SANS of M$_{75-25}$.} Parameters $A_1$ (pink squares) and $A_2$ (orange diamonds) from SANS fits as a function of temperature.}
\label{fig:M75_A1A2}
\end{figure}

\begin{figure}[!htb]
\centering
\includegraphics[scale=0.6]{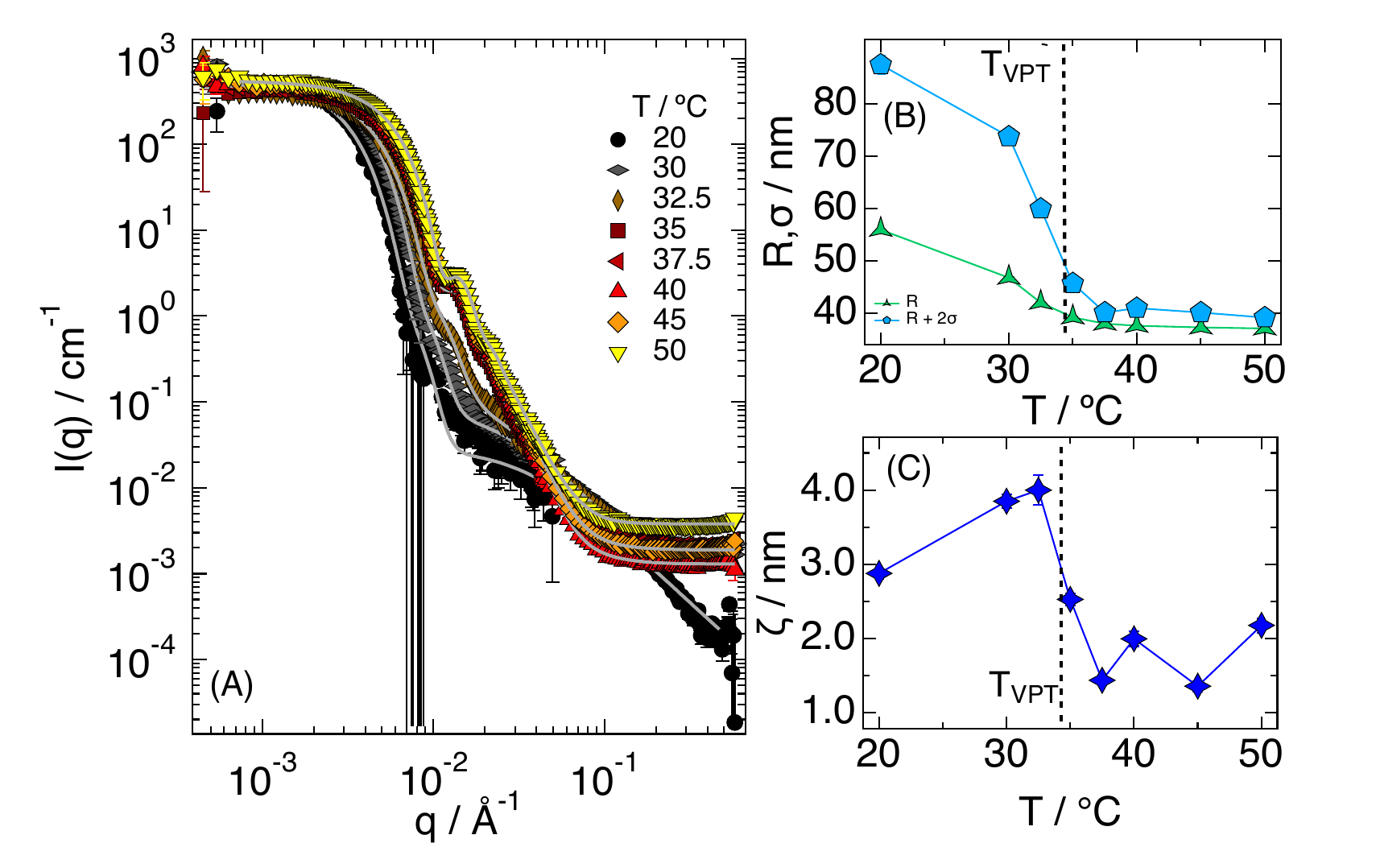}
\caption{\small \textbf{SANS of M$_{100-0}$.} Left: SANS scattering intensities $I(q)$ as a function of temperature. Lines represent fits with the fuzzy sphere model. Top-right: particle core radius ($R$) and total radius ($R + 2\sigma$) from SANS fits as a function of temperature. Bottom-right: correlation length of the polymer mesh ($\zeta$). $T_{VPT}$ (dashed line) indicates the VPTT from DLS analysis.}
\label{fig:M100}
\end{figure}

\begin{figure}[!htb]
\centering
\includegraphics[scale=0.4]{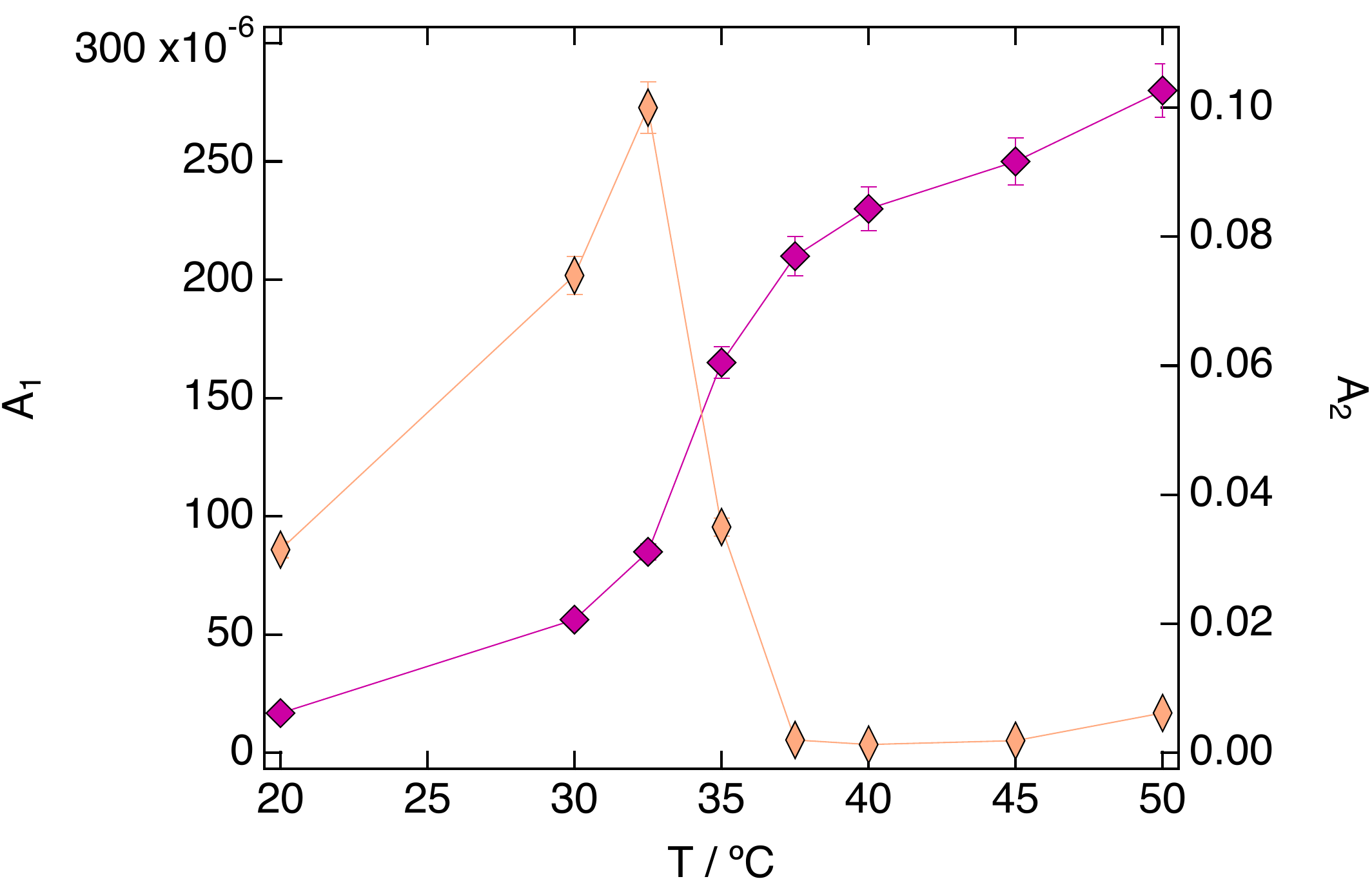}
\caption{\small \textbf{SANS of M$_{100-0}$.} Parameters $A_1$ (pink squares) and $A_2$ (orange diamonds) from SANS fits as a function of temperature.}
\label{fig:M100_A1A2}
\end{figure}

\begin{figure}[!htb]
\centering
\includegraphics[scale=0.6]{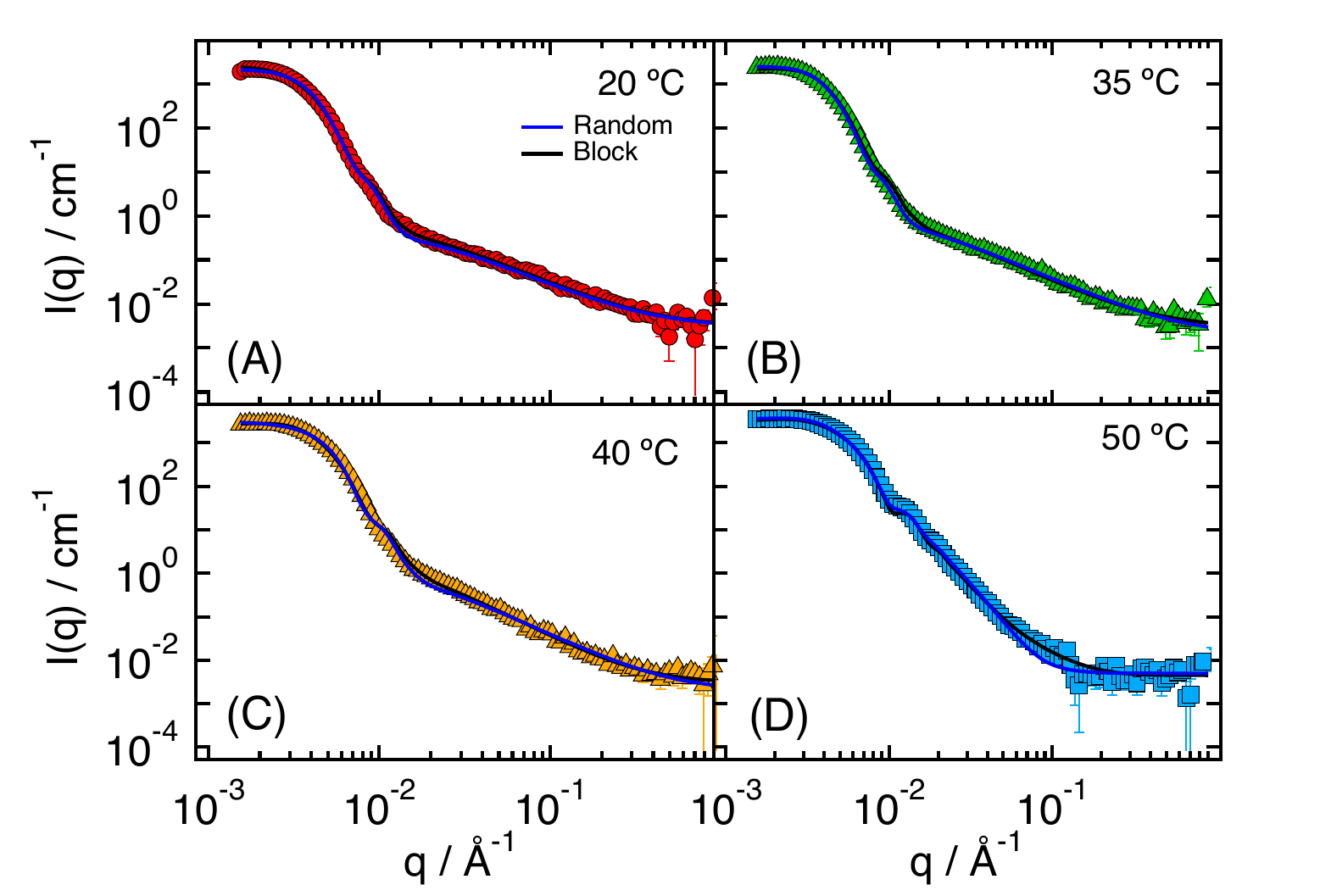}
\caption{\small \textbf{Comparison between experimental and numerical form factors.}  Experimental data (points) are measured for M$_{25-75}$ microgels (P(H-NIPAM-\textit{co}-H-NIPMAM)) at different temperatures. Black lines are numerical form factors obtained with the "\textit{block}" model while blue lines are numerical form factors obtained with the "\textit{random}" model (see Materials and Methods for details on the comparison procedure).}
\label{fig:M0}
\end{figure}

\begin{figure}[!htb]
\centering
\includegraphics[scale=0.6]{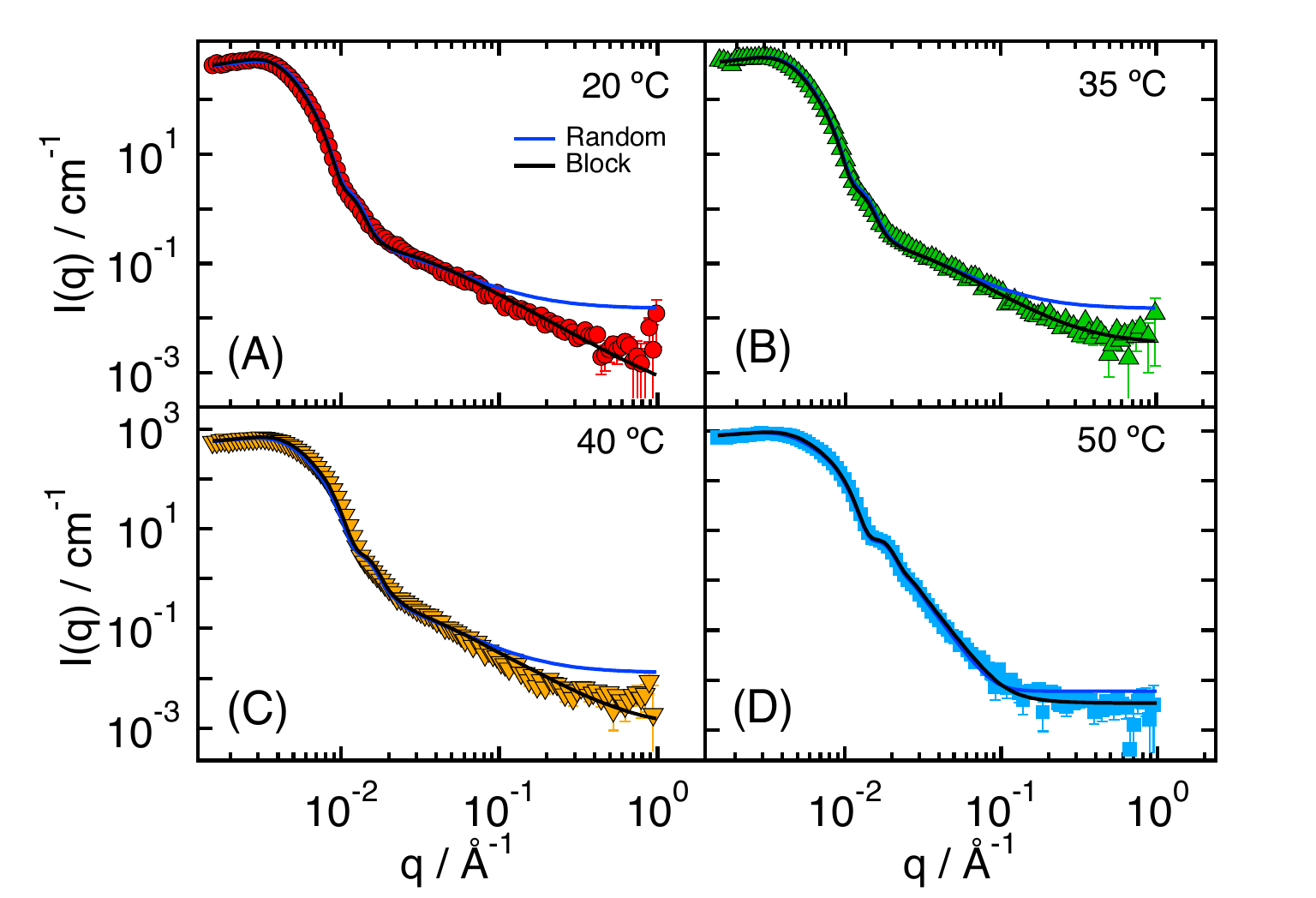}
\caption{\small \textbf{Comparison between experimental and numerical form factors.}  Experimental data (points) are measured for partially deuterated M$_{25-75}$ microgels (P(D-NIPAM-\textit{co}-H-NIPMAM)) at different temperatures. Black lines are numerical form factors obtained with the "\textit{block}" model while blue lines are numerical form factors obtained with the "\textit{random}" model (see Materials and Methods for details on the comparison procedure).}
\label{fig:M0D}
\end{figure}

\begin{figure}[!htb]
\centering
\includegraphics[scale=0.35]{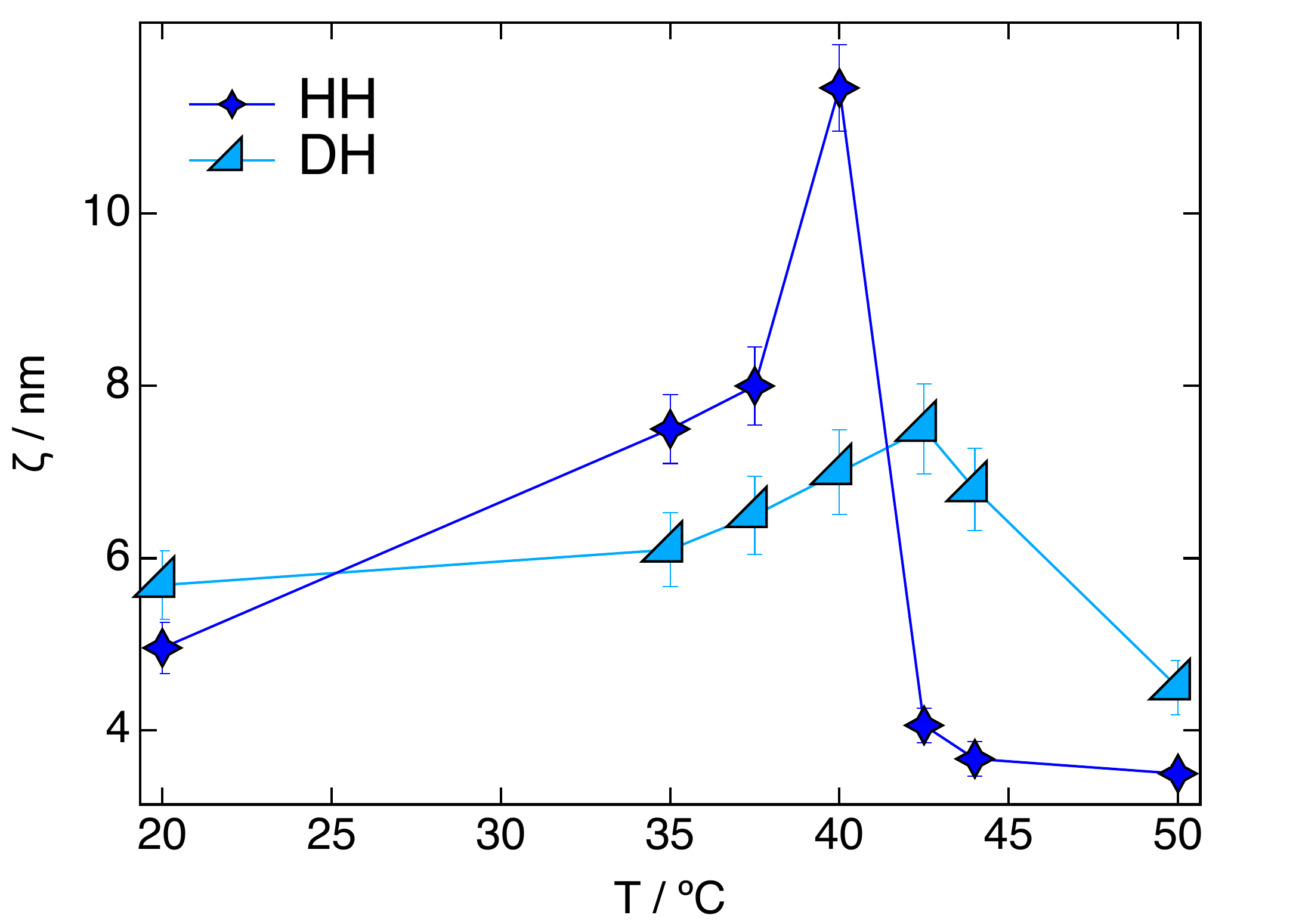}
\caption{\small \textbf{Correlation length of the polymer mesh ($\zeta$) for M$_{25-75}$ microgels containing H-NIPAM (HH) or D-NIPAM (DH).}}
\label{fig:Xi_HH_DH}
\end{figure}

\end{document}